\documentclass[prx,twocolumn,longbibliography,superscriptaddress,preprintnumbers]{revtex4-2}
\usepackage[colorlinks,bookmarks=true,citecolor=blue,linkcolor=blue,urlcolor=blue]{hyperref}
\usepackage{dcolumn,graphicx,amsfonts,amsthm,bm,color,appendix,float,tabularx}
\usepackage{braket}
\usepackage[normalem]{ulem}
\usepackage[version=4]{mhchem}
\usepackage{siunitx}
\usepackage{amsmath}
\usepackage{amssymb}
\usepackage{CJKutf8}
\usepackage{multirow}
\usepackage{booktabs}

\usepackage[dvipsnames]{xcolor}
\definecolor{pal0}{rgb}{0.8941, 0.102 , 0.1098}
\definecolor{pal1}{rgb}{0.2157, 0.4941, 0.7216}
\definecolor{pal2}{rgb}{0.302 , 0.6863, 0.2902}
\definecolor{pal3}{rgb}{0.5961, 0.3059, 0.6392}
\definecolor{pal4}{rgb}{1.    , 0.498 , 0.    }

\newcommand{\n}[1]{\left| #1 \right|}
\let\oldv\v
\renewcommand{\v}[1]{\boldsymbol{#1}}
\renewcommand{\Im}{\operatorname{Im}}

\newcommand{\Z}{\mathbb{Z}}

\newcommand{\PRLsec}[1]{\emph{#1---}}

\begin{document}

\title{
Topological constraint on crystalline current
}

\author{Tomohiro Soejima (\begin{CJK*}{UTF8}{bsmi}副島智大\end{CJK*})}
\affiliation{Department of Physics, Harvard University, Cambridge, MA 02138, USA}

\author{Junkai Dong (\begin{CJK*}{UTF8}{bsmi}董焌\end{CJK*}\begin{CJK*}{UTF8}{gbsn}锴\end{CJK*})}
\affiliation{Department of Physics, Harvard University, Cambridge, MA 02138, USA}

\author{Ophelia Evelyn Sommer}
\affiliation{Department of Physics, Harvard University, Cambridge, MA 02138, USA}

\author{Daniel E. Parker}
\affiliation{Department of Physics, University of California at San Diego, La Jolla, California 92093, USA}

\author{Ashvin Vishwanath}
\affiliation{Department of Physics, Harvard University, Cambridge, MA 02138, USA}

\begin{abstract}
How much current does a sliding electron crystal carry? The answer to this simple question has important implications for the dynamics of the crystal, such as the frequency of its cyclotron motion, and its phonon spectrum.
In this work we introduce a precise definition of a sliding crystal and compute the corresponding current $\v{j}_c$ for topological electron crystals in the presence of magnetic field. Our result is fully non-perturbative, does not rely on Galilean invariance, and applies equally to Wigner crystals and (anomalous) Hall crystals. In terms of the electron density $\rho$ and magnetic flux density $\phi$, we find that $\v{j}_c = e(\rho-C\phi)\v{v}$. Surprisingly, the current receives a contribution from the many-body Chern number $C$ of the crystal. When $\rho = C\phi$, sliding crystals therefore carry \textit{zero} current.
The crystalline current fixes the Lorentz force felt by the sliding crystal and the dispersion of low-energy phonons of such crystals.
This gives us a simple counting rule for the number of gapless phonons: if a sliding crystal carries nonzero current in a magnetic field, there is a single gapless mode, while otherwise there are two gapless modes. The constraint can also be understood from anomaly-matching of emanant discrete translation symmetries --- an idea that is also applicable to the dispersion of skyrmion crystals.
Our results lead to novel experimental implications and invite further conceptual developments for electron crystals. 
\end{abstract}

\maketitle

Recent advances in two-dimensional systems have rekindled interest in electronic crystals. 
Dating back to Wigner's original proposal in 1934~\cite{wigner_interaction_1934}, electron crystallization is one of the oldest strongly-correlated electronic phenomena.
Experimentally, such crystals were first observed in strong magnetic fields~\cite{lozovik_crystallization_1975,andrei_observation_1988,goldman1990evidence, jiang1990quantum,santos_observation_1992}, on the surface of liquid helium~\cite{grimes_evidence_1979}, and more recently visualized in two-dimensional heterostructures~\cite{tsui_direct_2023,xiang_quantum_2024}.

Topologically distinct from Wigner crystals, \textit{Hall crystals} are electron crystals that have nonzero quantized Hall conductance when pinned~\cite{kivelson_cooperative_1986,halperin_compatibility_1986,kivelson_cooperative_1987,tesanovic_hall_1989}. 
They were analyzed in detail in a seminal paper by Te\oldv{s}anovi\'{c} \textit{et al.} in 1989~\cite{tesanovic_hall_1989}. 
Recently, Hall crystals at zero magnetic field --- dubbed ``anomalous Hall crystals'' --- were proposed by some of the authors and other groups~\cite{AHC_Yahui,AHC1}, inspired by recent experiments on rhombohedral graphene heterostructures~\cite{lu2024fractional}.
As electron crystals are gapless, one expects that topology shapes their low-energy collective modes. In this work, we show that topology provides non-perturbative constraints on the crystalline current carried by a sliding electron crystal.

Informally, the crystalline current $\v{j}_c$ is the current carried by a crystal sliding at velocity $\v{v}$.  For instance, a Wigner crystal with density $\rho$ has $\v{j}_c = e \rho \v{v}$, while a crystal of neutral excitons has $\v{j}_c = 0$. Our main result is that a Chern-$C$ electron crystal has crystalline current
\begin{equation}
\label{eq:crystalline_current_formula}
    \v{j}_c = e(\rho - C \phi) \v{v},
\end{equation}
where $\rho$ is electron density and $\phi$ is flux density. We make five comments, evidenced fully below.
(I) This result is topological; the quantity $s = A_{\mathrm{uc}}(\rho - C \phi)$ is the ``bound charge" topological invariant in the Widom-St\oldv{r}eda formula~\cite{widom1982thermodynamic, streda1982theory} $\rho = C \phi + s/A_{\mathrm{uc}}$.
(II) Historically, Eq.~\eqref{eq:crystalline_current_formula} reaches back to the 1980s, with precedents in Refs.~\cite{tesanovic_hall_1989} and ~\cite{kunz1986quantized} at the mean-field or single-particle level. Our derivation of Eq.~\eqref{eq:crystalline_current_formula} is a many-body, non-perturbative result. It applies at both finite and zero field, and even holds for fractional $C$.
(III) 
Sliding crystals feel a Lorentz force
\begin{equation}
\label{eq:crystal_lorentz_force}
    \beta = eB(\rho-C\phi)\rho^{-1}.
\end{equation}
This also controls the number of gapless phonon modes: when $\beta=0$, there are two such modes, and otherwise there is one mode.
(IV)
Anomaly matching of discrete magnetic translation symmetry between the UV and IR fixes the form of Eq.~\eqref{eq:crystalline_current_formula} up to an integer multiple of $e\phi \v{v}$.
(V)
Experimentally, the coupling between electric fields and electron crystals is proportional to $s$. This sharply modifies electromagnetic responses, especially in the case of full Hall crystals where $s=0$.
 
\PRLsec{Magnetic Translations on a Torus}
We consider a two-dimensional spatial torus
with a perpendicular magnetic field $B$ and continuous magnetic translation symmetry. Our results will hold for any torus, but for simplicity we use periods $\v L_1 = L_x \hat{\v e}_x$, $\v L_2 = L_y \hat{\v e}_y$ with total magnetic flux $N_\Phi = eB L_x L_y/ 2\pi \hbar \in \Z$
~\cite{dirac1931quantised}. Below we set $\hbar = 1$, system size $A = L_x L_y$, and adopt Landau gauge $\v A = (-By, 0)$. The action of continuous magnetic translation by $\v{u} = (u_x,u_y)$ on single-particle wavefunctions is
\begin{equation}
    \hat{T}_{\v u}\psi(x,y) = e^{ieB xu_y}\psi(x-u_x,y-u_y).
\end{equation}
These do not commute but instead satisfy $T_{\v{u}_1}T_{\v{u}_2} = e^{i eB \v{u}_1 \wedge \v{u}_2} T_{\v{u}_2} T_{\v{u}_1}$, where $\v{a} \wedge \v{b} = a_x b_y - a_y b_x$.

Different Hilbert spaces $\mathcal{H}_{\v \Phi}$ on the torus are characterized by the flux $\v{\Phi}$ through the handles of the torus, i.e. the boundary conditions $T_{\v L_j}\psi(x,y) = e^{i\Phi_j}\psi(x,y)$ on single-particle states. Crucially, the non-commutative nature of magnetic translations implies their action changes the boundary conditions: if $\psi\in \mathcal{H}_{\v \Phi}$, then $\psi'=T_{\v u}\psi \in \mathcal{H}_{\v \Phi+\Delta \v \Phi}$, where 
\begin{equation}
    \Delta\v{\Phi} = eB(L_xu_y, -L_yu_x).
    \label{eq:mag_translation_action_on_flux}
\end{equation}

\begin{figure}
    \centering
    \includegraphics[width=\linewidth]{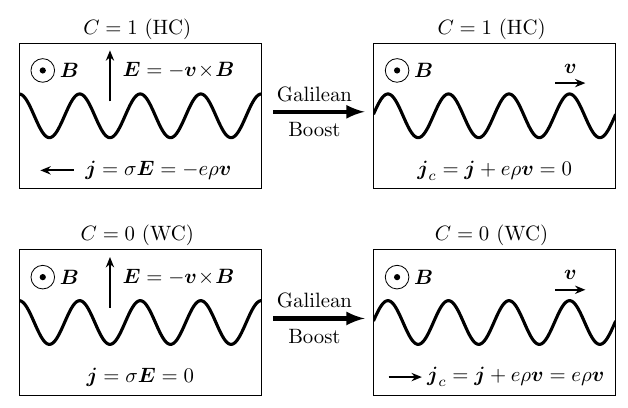}
    \caption{Demonstration of the crystalline current formula $\v{j}_c = e(\rho-C\phi)\v{v}$ when the system is Galilean invariant. The crystalline current is generated by a sliding crystal state Eq.~(\ref{eq:sliding_crystal}). To compute the current carried by the crystal, we first compute the current generated by the electric field $-\v{v}\times\v{B}$ when the crystal is stationary. Then, we perform a Galilean boost to the system. The Galilean boost generates an electric field that cancels the applied electric field, and the electronic state is exactly given by Eq.~(\ref{eq:sliding_crystal}). The crystalline current is the sum of the current induced by the electric field and the current generated by the Galilean boost. We show the cases of a full Hall crystal (HC) $\rho  = C\phi$ and a Wigner crystal (WC).
    }
    \label{fig:Galilean_cartoon}
\end{figure}

\PRLsec{Crystalline current}Let the many-body Hamiltonian for $N_e$ electrons on the torus be
\begin{equation}
    \hat{H} = \sum_{j=1}^{N_e} \hat{h}(\v p_j-e\v A) + \frac{1}{2}\sum_{i,j= 1}^{N_e} \hat{V}(\v r_i - \v r_j),
\end{equation}
where $\hat{h}(\v{p})$ is a single-particle Hamiltonian and $\hat{V}$ is a two-body interaction. $\hat{H}$ obeys continuous magnetic translation symmetry: $[\hat{H}, \hat{T}_{\v{u}}] = 0$.

We assume that the ground state of $\hat{H}$ is a \textit{crystalline insulator}.
Explicitly, consider a small pinning field $\hat{H}_{\mathrm{pin}} = \lambda U(\v{r})$ that breaks continuous translation symmetry.
We require that if we take the thermodynamic limit $A \to \infty$, then $\hat{H} + \hat{H}_{\mathrm{pin}}$ has a unique crystalline ground state at any sufficiently small finite $\lambda$, which we can then take to zero~\cite{Brauner_2024}\footnote{As is the case for all pinning fields for symmetry broken states, the effect of this potential on the ground state is highly non-linear}. We further impose that said ground state is an insulator: its many-body gap $\Delta(\lambda, A)$ converges to nonzero $\Delta(\lambda)$ as $A\to\infty$ at $\lambda >0$. A  metal coexisting with crystalline order, by contrast, would remain gapless even when pinned.

We now define a sliding crystal.
Consider $\ket{\v{u};\v{\Phi};\lambda}$, which is the unique ground state of the Hamiltonian with a displaced pinning potential $\hat{H}+\lambda U(\v r-\v u)$~\footnote{As before, we take the thermodynamic limit before taking the $\lambda \to 0$ limit.}.
A \textit{sliding crystal} with velocity $\v{v}$ is defined to be a trajectory in this manifold:
\begin{equation}
    \ket{\Psi(t)} = \ket{\v u(t);\v{\Phi}}, \quad \v u(t) = \v{v} t,
    \label{eq:sliding_crystal}
\end{equation}
where we suppressed the dependence on $\lambda$.
This definition is physically motivated as follows. Consider subjecting a crystal to the time dependent Hamiltonian $\hat{H}+\lambda U(\v r-\v u(t))$ for small but nonzero $\lambda$. Since the state is gapped in the thermodynamic limit, we can take $\v{v}$ sufficiently small so that time evolution is well-approximated by $\ket{\Psi(t)}$ via the adiabatic approximation.
In other words, Eq.~\eqref{eq:sliding_crystal} corresponds to a crystal that exactly tracks a moving external potential.\footnote{{
When does a crystal follow the dragging potential? Let $\Delta$ be the gap induced by the dragging potential, $v$ the velocity of the crystal, and $a$ the lattice constant of the crystal. The adiabatic approximation holds when $\omega = v / a \ll \Delta/\hbar \implies v \ll a\Delta / \hbar  $.
For example, a Wigner crystal with a lattice constant of $10$ nm and a pinning gap of $1$ meV has $v = 1.5 \times 10^4$ m/s, showing that a very fast motion is possible.
In the absence of an explicit dragging potential, the role of the dragging potential will be played by the self-consistent field generated by the crystal itself, and its strength will determine the regime of validity of our formula.
}}

The \textit{crystalline current} is defined to be the current carried by $\ket{\Psi(t)}.$ Without loss of generality, consider $\v{v}$ in the $x$ direction.
The current density in the $x$ direction is then:
\begin{equation}
\begin{aligned}
\label{eq:crystalline_current_definition}
 j_c = \frac{e}{L_xL_y}\frac{dP_x}{dt}, 
 \end{aligned}
\end{equation}
where
\begin{equation}
    P_x(t) = \frac{L_x}{2\pi}\,\textrm{Im}\,\log\braket{\Psi(t)|e^{2\pi i \hat{x}/L_x}|\Psi(t)}, \quad \hat{x} = \sum_{j=1}^{N_e} \hat{x}_j
\end{equation}
is the many-body polarization~\cite{resta1998quantum}. 

\PRLsec{Derivation of Eq.~\eqref{eq:crystalline_current_formula}} As the crystalline current should be independent of the boundary condition $\v{\Phi}$, we set $\v{\Phi}=0$ without loss of generality~\cite{HARUKI2018}. Using Eq.~\eqref{eq:mag_translation_action_on_flux}, the state $\ket{\Psi(t)}$ can be written as the ground state of $\hat{H}_{\v u=0}$ in a different flux sector after magnetic translation:
\begin{equation}
\label{eq:translate_to_zero_COM}
    \ket{\Psi(t)} 
    = \hat{T}_{\v{ u}} \ket{0,-\Delta \v{\Phi}},
\end{equation}
where $\v{u}$ and $\Delta \v{\Phi} = eB(L_x u_y(t), -L_y u_x(t))$ depend implicitly on time. The time-dependence splits into two parts: the action of magnetic translations $T_{\v{u}}$, and the action of changing flux, which gives an electromotive force that reveals the Hall response of the system. 

Using the identity 
\begin{equation}
\label{eq:magnetic_translation_conjugation}
    \hat{T}^\dagger_{\v{u}} e^{2\pi i \hat{x} / L_x} \hat{T}_{\v{u}}
    = e^{2\pi i (\hat{x} + N_e u_x) / L_x},
\end{equation}
the polarization becomes $P_x(t) = N_e u_x(t) + \tilde{P}_x(\Delta \v{\Phi}(t))$ where 
\begin{equation}
    \tilde{P}_x(\Delta \v{\Phi}) =\frac{L_x}{2\pi}\Im \log\braket{0, -\Delta \v{\Phi}|e^{2\pi i \hat{x}/L_x}|0, -\Delta \v{\Phi}}.
\end{equation}
As this is the polarization of a time-independent Hamiltonian $\hat{H}_{\v u=0}$, changing flux reflects the Hall response~\cite{ManyBodyCN}
\begin{equation}
    \frac{d \tilde{P}_x}{d\Phi_y} = -\frac{C}{2\pi}L_x,
    \label{eq:many_body_chern_number_from_polarization}
\end{equation}
where $C$ is the many-body Chern number~\footnote{We use the convention such that for $eB>0$ the lowest Landau level has $C=1$.}. Hence
\begin{equation}
    \begin{aligned}
        \frac{d P_x}{dt}
        = N_e \frac{du_x}{dt} + \frac{d\tilde{P}_x}{d\Phi_y} \frac{d\Phi_y}{du_x}\frac{du_x}{dt}
        = \left(N_e - C \frac{BL_xL_y}{2\pi} \right) v_x.
    \end{aligned}
\end{equation}
From the definition Eq.~\eqref{eq:crystalline_current_definition}, the average current density passing through a spatial slice along the $y$ direction is therefore
\begin{equation}
        j_x = e(\rho-C\phi)v_x,
        \label{eq:topological_formula_for_current}
\end{equation}
where $\rho=N_e/L_xL_y$ is the charge density, and $\phi=N_{\Phi}/L_xL_y$ is the flux density. One can compute that the current in the $y$ direction is zero, yielding Eq.~\eqref{eq:crystalline_current_formula}. This is the main result of this paper: the current carried by a sliding crystal depends on the Chern number.

We can simplify Eq.~(\ref{eq:topological_formula_for_current}) by the Widom-St\oldv{r}eda formula~\cite{widom1982thermodynamic, streda1982theory}:
\begin{equation}
    \rho = C\phi + \frac{s}{A_{uc}},
\end{equation}
where $A_{uc}$ is the area of the unit cell, and $s$ is the ``bound-charge" invariant" that counts the number of electrons per unit cell in the limit of $B\to 0$. 
Putting this into Eq.~\eqref{eq:crystalline_current_formula} gives
\begin{equation}
    \v{j}_c = \frac{s e }{A_{uc}}\v{v}.
    \label{eq:bound_charge_formulation}
\end{equation}
So
when the pinning potential is moved, only the bound charge is dragged along with it \footnote{This argument was used to predict our result(Eq.~\ref{eq:bound_charge_formulation}) in Ref.~\cite{tesanovic_hall_1989}.}.
In Ref.~\cite{tesanovic_hall_1989}, crystalline states with nonzero $C$ and zero $s$ or nonzero $s$ were named ``full'' Hall crystals or ``partial'' Hall crystals respectively. We see that sliding {\em full} Hall crystals carry zero current.

\PRLsec{Galilean invariance} When the Hamiltonian is Galilean invariant, Eq.~\eqref{eq:crystalline_current_formula} follows directly from macroscopic properties of the crystal.
Consider an initial configuration with a pinning potential $\lambda U(\v{r})$ and an electric field $\v{E} = -\v{v}\times \v{B}$ (Fig.~\ref{fig:Galilean_cartoon}). This generates a Hall current density $\v{j} =  \sigma  \v{E} = -eC \phi \v{v}$, where $\sigma$ is the conductivity tensor. Let us now act with a Galilean boost with velocity $\v{v}$. The potential is now $\lambda U(\v{r} - \v{v}t)$ and $\v{E} = 0$, which is precisely what generates a sliding motion. The Galilean boost generates an additional current equal to $e\rho \v{v}$, resulting in $\v{j}_c = e(\rho - C \phi)\v{v} $, which agrees with Eq.~\eqref{eq:crystalline_current_formula}.

\PRLsec{Examples}Let us first consider crystalline states in the lowest Landau level. The Hamiltonian has a particle-hole symmetry that acts as
\begin{equation}
    \rho \to \phi-\rho, \quad C \to 1-C, \quad \v{j} \to -\v{j},
\end{equation}
where $C$ is the Chern number of the ground state~\cite{DanArovasQHENotes}.
The current formula Eq.~\eqref{eq:crystalline_current_formula} is compatible with the transformation. In particular, a Wigner crystal of electrons with $\rho = \rho_0, C = 0, \v{j} = e\rho_0 \v{v}$ gets transformed to a Wigner crystal of holes  with the opposite and equal current under this transformation: $\rho = \phi - \rho_0, C = 1, \v{j} = e(\phi - \rho_0 - \phi) \v{v} = -e\rho_0 \v{v}$~\footnote{We sincerely thank Patrick J. Ledwith for this simple but illuminating argument.}.

A second example is Wigner crystals in high Landau levels. Consider the Landau level filling $\nu = \rho/\phi = m + \epsilon$, where $m$ is some large integer. Let us assume the first $m$ Landau levels are fully filled so that  the Chern number is $C=m$, and the remaining $\epsilon$ electrons form a Wigner crystal. Then the crystalline current is $\v{j}_c \propto \epsilon$, meaning the filled Landau levels are frozen~\footnote{In fact, our proof actually did not rely on the states having nonzero crystalline order parameter. Therefore, our derivation showed that fully filled Landau levels with $\nu \in \mathbb{Z}$ carry zero current under moving pinning field, a trivial result given that the state is not affected by pinning.}.

As discussed earlier, full Hall crystals carry zero current. This can be understood as follows~\cite{tesanovic_hall_1989, HallCrystalTDHF}. Full Hall crystals have  $\rho = C\phi$, corresponding to filling $\nu=C$ Landau levels. They can thus be understood as crystals of particle-hole pairs, or excitons, between different Landau levels, on top of fully filled Landau levels. Since  excitons carry no charge, the sliding full Hall crystal does not generate electric current.

\PRLsec{Implication for phonons}The crystalline current is closely linked to the collective mode excitations above the electronic ground state, i.e. the phonon spectrum. Indeed, the phonon mode at momentum $\v{k}=0$ is a sliding crystal, whose current couples to the magnetic field to generate Lorentz force. We will see that Eq.~\eqref{eq:crystalline_current_formula} directly determines the low-energy properties of phonons, including the number of gapless modes.

To start, we can use the value of the crystalline current to fix the low-energy effective action for the phonon, whose kinetic part, up to second order time derivatives, is
\begin{equation}
    S_k = \frac{\rho}{2}\int d^2 \v r dt \left(m \dot{\v{u}}^2 + \beta  \dot{\v{u}}\times \v{u} \right),
    \label{eq:phonon_action}
\end{equation}
where $\v{u}(\v{r},t)$ is the crystal displacement field, $\dot{\v{u}} = d \v{u}/dt$,
$m$ is the effective mass, and $\beta$ is a Lorentz term.
Hence the equation of motion for the uniformly sliding crystal is
\begin{equation}
    m \ddot{\v{u}} = \beta \dot{\v{u}} \times \hat{z},
\end{equation}
where $\hat{z}$ is the unit vector in the $z$ direction. On the other hand, microscopically, the crystal feels the Lorentz force $\v{F}_L$ due to the current it generates:
\begin{equation}
    m\ddot{\v{u}} = \frac{\v F_{L}}{N_e} = \rho^{-1}\v j\times \v B = eB\left(1-\frac{C\phi}{\rho}\right)\dot{\v{u}} \times \hat{z}.
\end{equation}
Matching these two expressions, we find
\begin{equation}
    \beta=eB(1-C\phi/\rho)=seB/\rho A_{uc},
    \label{eq:beta_for_phonons}
\end{equation}
which implies Eq.~\eqref{eq:crystal_lorentz_force}.

We can understand the effect of $\beta$ on the $\v{k} \neq 0$ phonon spectrum by restoring elastic terms that depend on partial derivatives $\partial \v{u}$ in Eq.~\eqref{eq:phonon_action}. The resulting action is equivalent to that of a Wigner crystal in a magnetic field, where $\beta$ is replaced by $eB$, which has been well-studied~\cite{Fukuyama_magnetic_1975, Bonsall-Maradudin}. In the absence of $\beta$, there are two gapless modes, corresponding to transverse and longitudinal motions. Nonzero $\beta$ mixes these two modes, resulting in frequencies $\omega=0$ and $\omega=\beta/m$~\footnote{This may differ from the cyclotron gap of the entire system, $\omega_c = eB/m_e$, where $m_e$ is the electron mass, as dictated by Kohn's theorem~\cite{kohn1961cyclotron}. This is because $\omega_c$ corresponds to the motion of all electrons at the same time, which differ from the motion of bound-charges considered here.}.
Nonzero $\beta$ therefore changes the number of gapless modes, resulting in a so-called ``anomalous dispersion''. 
The single gapless phonon will have a dispersion proportional to the product of the dispersions of the two gapless modes before mixing, which depend on the interactions~\cite{Fukuyama_magnetic_1975}.

The nontrivial topology of Hall crystals arises from Berry curvature, which is an analogue of a magnetic field in momentum space. 
While magnetic fields cause cyclotron motion, Berry curvature alone does not. From Eq.~\eqref{eq:beta_for_phonons}, the Lorentz force vanishes either when the crystal carries no current ($s=0$), or when there is no magnetic field ($B=0$). The former corresponds to full Hall crystals~\cite{tesanovic_hall_1989,HallCrystalTDHF}, and the latter corresponds to ``anomalous Hall crystals''~\cite{AHC_Yahui,AHC1}. We pointed out the absence of $\beta$ for the latter in Ref.~\cite{AHC4}. The number of zero modes for various topological crystals are summarized in Table~\ref{tab:phonon_gap_of_crystals}. This compliments the existing understanding of Goldstone mode count for internal symmetries~\cite{watanabe2020counting}.

\PRLsec{Higher Frequency Corrections}
At higher order in frequency, terms ignored in Eq.~\eqref{eq:phonon_action}, such as the effect of momentum-space Berry curvature, modify the cyclotron dynamics, in particular changing the gap size from $\omega = \beta/m$.
The momentum space Berry curvature is defined as $\Omega^\mathrm{pho}=- iN_e\bra{\partial_{\v p}\Psi_{\v p}}\times\ket{\partial_{\v p}\Psi_{\v p}}$ where $\ket{\Psi_{\v p}}$ is the low energy state with momentum $\v{p}$~\footnote{The prescription for preparing this state was discussed in Ref.~\cite{Zeng_sliding}, including the proper method for gauge-fixing.}. 
In analogy with a single particle moving in the presence of a magnetic field and Berry curvature~\cite{xiao2010berry, chang1996berry, chang1995berry}, the first order phase space action of the center of mass motion is then captured by 
\begin{equation}
     S[\v u,\v p]=\int d t ~\v p\cdot \dot{\v u}+\frac{N_e\beta}{2}\dot{\v u}\times \v u+\frac{\Omega^\mathrm{pho}}{2N_e}\dot{\v p}\times \v p-\frac{\v p^2}{2mN_e},\label{eq:action_phasespace}
\end{equation}
where we note $\beta$ replaces the role of the bare magnetic field.
Integrating out $\v{p}$ and taking $\omega \ll 1/m\Omega^\mathrm{pho}$, we find an effective action
\begin{equation}
    S_k = \frac{\rho}{2}\int d^2 \v r dt \left(m \dot{\v{u}}^2 + \beta  \dot{\v{u}}\times \v{u} +m^2\Omega^\mathrm{pho}\ddot{\v u}\times \dot{\v u}+\dots\right). 
    \label{eq:phononBC_action}
\end{equation}
As $\Omega^\mathrm{pho}$ shows up at higher order in time derivatives, it is subdominant to $\beta$ at low frequency. However, the cyclotron motion appears at higher frequency, and ignoring corrections beyond Eq.~\eqref{eq:action_phasespace} it is modified to $\omega_c=\beta/[m(1+\beta\Omega^\mathrm{pho})]$~\footnote{The cyclotron frequency satisfies $\omega_c \ll 1/m\Omega^\mathrm{Pho}$ provided $\beta\Omega \ll 1$. Similar discussion on the role of $\Omega^\mathrm{Pho}$ can be found in Ref.~\cite{joy2025marchmeeting}}.
Observations of the cyclotron frequency alone are therefore \emph{not} a direct measurement of $\beta$.
The apparent divergence of $\omega_c$ at $\beta\Omega^\mathrm{pho} = -1$ is unphysical, since it goes beyond the range of validity of low-energy effective action.

\PRLsec{$\beta$ as a Berry phase} The $\beta$ term can also be interpreted as a Berry phase term. This allows us to compute $\beta$ by matching the Berry phase generated by this term and the Berry phase corresponding to an adiabatic trajectory of crystalline ground states.

First, consider an adiabatic sliding motion of the crystal under the action in Eq.~\eqref{eq:phonon_action}, which moves according to the trajectory $\v u(t)$, where $\v u(T) = \v u(0)$. Then, the $\beta$ term contributes a Berry phase:
\begin{equation}
    \varphi_\beta = N_e \beta A_\circ,
    \label{eq:berryaction}
\end{equation}
where $A_\circ$ is the signed area enclosed by the loop $\v u(t)$. 

Microscopically, such a motion is represented by the trajectory of states $\ket{\Psi(t)} = \ket{\v u(t);\v\Phi=0}$. 
The Berry phase associated with these states must match Eq.~(\ref{eq:berryaction}):
\begin{equation}
    e^{i\varphi_\beta} = e^{i\int_0^T A(t) dt},\quad A(t)= -i\braket{\Psi(t)|\partial_t \Psi(t)}.
\end{equation}

Similar to the $\v{j}_c$ calculation, the time dependence of the state has contributions from
both the time-dependent magnetic translation and the time-dependent flux sectors (see App.~\ref{app:berry_phase_for_crystals}). The resulting Berry phase is
\begin{equation}
    \varphi_\beta = N_e\,eB (1- C\phi / \rho),
    \label{eq:berryphase}
\end{equation}
which agrees with Eq.~(\ref{eq:beta_for_phonons}).

We note that this Berry phase encodes non-commutativity of translation operators, just as the Aharonov-Bohm phase from magnetic fields gives rise to the magnetic translation algebra. Curiously, summing over the microscopic Aharonov-Bohm phase for all electrons seems to imply $\varphi_\beta = eN_e B A$. Yet this is only true for crystalline states with $C=0$. In fact, full Hall crystals with $\rho = C\phi$ have \textit{commuting} translations, despite their non-commuting microscopic translations. We resolve this seeming mismatch in the next section.

\PRLsec{Anomaly of Emanant Translation Symmetry}
To understand the commutativity of translation operators in the IR, we will carefully analyze the constraint imposed by the microscopic theory.
Importantly, we find that the algebra of \textit{discrete} translation symmetries need to be matched between UV and IR.

Recall that the action of magnetic translation changes the flux sector as Eq.~\eqref{eq:mag_translation_action_on_flux}. Therefore, infinitesimal magnetic translation does not preserve the Hilbert space~\footnote{This breaking of continuous translation symmetries has been discussed in Refs.~\cite{fischler1979quark, metlitski2007schwinger, seiberg, seiberg2025anomalouscontinuoustranslations} as well as Appendix G of Ref.~\cite{geraedts2016half}.}. The minimal translations that preserve the boundary conditions satisfy $\Delta\v{\Phi} = (2\pi, 0)$ and $\Delta\v{\Phi} = (0, 2\pi)$, which corresponds to
\begin{equation}
    \v{b}_1 = \frac{2\pi}{eBL_y}(1, 0), \quad  \v{b}_2 = \frac{2\pi}{eBL_x}(0, 1).
\end{equation}
It is the algebra of these \textit{discrete} translations in the microscopic theory that constrains $\beta$ in the effective action, not those of continuous translations.
For a system of $N_e$ electrons, the magnetic translation algebra for these minimal translations are
\begin{equation}
    T_{\v{b}_1} T_{\v{b}_2} =  e^{iN_e eB \v{b}_1 \wedge \v{b}_2} T_{\v{b}_2} T_{\v{b}_1} = e^{i2\pi \frac{N_e}{N_\Phi}} T_{\v{b}_2} T_{\v{b}_1},
    \label{eq:anomaly_matching_constraint}
\end{equation}
whose non-commutativity is a manifestation of the UV mixed anomaly between two translation operators.
On the other hand, the Berry phase associated with a trajectory around a plaquette spanned by $\v{b}_1$ and $\v{b}_2$ is given by $N_e \beta (2\pi)^2 / (B^2 L_x L_y) = \beta  \rho / B\phi$.
We now perform ``anomaly matching'': we demand these two phases match between the UV and IR theories, which immediately implies
\begin{equation}
    \beta = eB(\rho - m \phi) / \rho \text{   with } m \in \mathbb{Z}.
    \label{eq:beta_from_emanant}
\end{equation}
Thus, anomaly matching fixes $\beta$ up to an offset $m$. By comparison with Eq.~\eqref{eq:beta_for_phonons}, we immediately see $m=C$. In particular, when $\nu \in \mathbb{Z}$, the IR translation operators can commute despite the magnetic translation algebra~\footnote{A similar anomaly-matching argument has been used to derive Luttinger's theorem for Fermi liquids~\cite{oshikawa2000topological, else2021non, cheng2023lieb}}.

Anomaly matching of discrete symmetries can arise in different physical contexts.
As discussed in Ref.~\cite{seiberg, seiberg2025anomalouscontinuoustranslations} and Appendix~\ref{app:skrymion}, the same consideration applies to the dynamics of skyrmions, and by extension, skyrmion crystals~\cite{nagaosa2013topological}, where the coefficient of the Berry phase term is similarly quantized according to the skyrmion winding number~\cite{stone1996magnus}.
These two examples provide an important lesson: discrete emanant symmetry can constrain the Berry phase term \textit{up to a topological invariant}, whose identity must be established microscopically.

\PRLsec{Generalization to Fractional Hall crystals} Our discussion so far assumed crystalline insulators with unique ground states under pinning, forcing the Chern number $C$ to be an integer. In fact, Eq.~\eqref{eq:crystalline_current_formula} generalizes easily to fractional Hall crystals with fractional $C$. For the Galilean invariant case, the argument above does not depend on whether $\sigma_{xy}$ takes a fractional value. The main challenge for a more rigorous proof is the bookkeeping required for multiple quasi-degenerate ground states,
which we address carefully in Appendix~\ref{app:fractional_hall_crystals}. We also find that fractional states modify the topological constraint Eq.~\eqref{eq:beta_from_emanant} imposed by emanant symmetry.


\begin{table}
    \centering
    \caption{Phonon properties of various topological crystals: Wigner crystals (WC), magnetic Wigner Crystals, anomalous Hall crystals (AHC), and full and partial Hall crystals (HC). The number of gapless phonon modes at $\v{k}=0$ is $2$ if $\beta=0$ and $1$ otherwise.}
    \label{tab:phonon_gap_of_crystals}
    \begin{ruledtabular}
    \begin{tabular}{lcccc}
       Crystal  & $s$ & $B$ & $\beta$ & Gapless Phonons\\ \hline
       WC  &  $\neq 0$ & 0 & 0 & 2\\
       $\n{B} > 0$ WC & $\neq 0$ & $\neq 0$ & $\neq 0$ & $1$\\
       AHC & $\neq 0$ & $0$ & $ 0$ & $2$\\
       Full HC & $0$ & $\neq 0$ & $ 0$ & $2$\\
       Partial HC & $\neq 0$ & $\neq 0$ & $\neq 0$ & $1$\\
    \end{tabular}
    \end{ruledtabular}
\end{table}

\PRLsec{Experimental implications} Our most striking prediction is the vanishing of the current $\v{j}_c$ for full Hall crystals.
In the $\rho-\phi$ phase diagram of topological crystals, measurements of the phonon spectrum can reveal the enhancement from one to two gapless modes exactly when $\rho = C\phi$. Full Hall crystals are further distinguished by their unusual electromagnetic response: as they carry zero current when sliding, electric fields do not couple to their center of mass motion. As we show in Appendix~\ref{app:hall_crystal_response}, this can be detected by longitudinal conductivity of a pinned crystal: to the lowest order in $\omega$, the conductivity scales as $\sigma^{xx}(\omega) \propto s^2 \omega$. At large voltage, we expect a large non-linearity in the response, resulting from depinning of the crystal~\cite{gruner1988dynamics,gruner2018density,fisher1998collective, brazovskii2004pinning, reichhardt2016depinning, patri2024extended}. 

We close by pointing out a natural place to look for full Hall crystals: bilayer quantum Hall systems at filling $\nu = \nu_t + \nu_b  \in \mathbb{Z}$, where $\nu_t$ and $\nu_b$ are filling of the top and bottom layers. When $\nu_t \ll \nu_b$, the system can be thought of as a low-density gas of interlayer excitons at density $\nu_t$. If these excitons crystallize, the resulting excitonic crystal is precisely a full Hall crystal. This could be readily detected by applying electric field to a single layer, evading the vanishing electric coupling. Depinning the crystal thusly would create perfect Coulomb drag currents $\v{j}_t = -\v{j}_b$ with $\v{j}_c = 0$, just as in excitonic condensates~\cite{nandi2012exciton, narozhny2016coulomb}.

\begin{acknowledgements}
We thank Taige Wang, Tianle Wang, and Mike Zaletel for related collaborations and useful insights. We thank Ethan Lake, Haruki Watanabe, Andy Millis, Yugo Onishi and especially Patrick J. Ledwith and Bert Halperin for fruitful discussions.
This research is funded in part by the
Gordon and Betty Moore Foundation’s EPiQS Initiative,
Grant GBMF8683 to T.S.; A.V., O.E.S. and J.D. were funded by NSF DMR-2220703.
AV is supported by the Simons Collaboration on Ultra-Quantum Matter, which is a grant from the Simons Foundation (651440, A.V.).
D.E.P. acknowledges startup funds from UC San Diego.
\end{acknowledgements}

\let\temp\oldv
\let\oldv\v
\let\v\temp

\bibliographystyle{unsrt}
\bibliography{references}

\begin{thebibliography}{10}

\bibitem{wigner_interaction_1934}
E.~Wigner.
\newblock On the {Interaction} of {Electrons} in {Metals}.
\newblock {\em Phys. Rev.}, 46(11):1002--1011, December 1934.
\newblock Publisher: American Physical Society.

\bibitem{lozovik_crystallization_1975}
Yu.~E. Lozovik and V.~I. Yudson.
\newblock Crystallization of a two-dimensional electron gas in a magnetic field.
\newblock {\em Soviet Journal of Experimental and Theoretical Physics Letters}, 22:11, July 1975.
\newblock ADS Bibcode: 1975JETPL..22...11L.

\bibitem{andrei_observation_1988}
E.~Y. Andrei, G.~Deville, D.~C. Glattli, F.~I.~B. Williams, E.~Paris, and B.~Etienne.
\newblock Observation of a {Magnetically} {Induced} {Wigner} {Solid}.
\newblock {\em Phys. Rev. Lett.}, 60(26):2765--2768, June 1988.
\newblock Publisher: American Physical Society.

\bibitem{goldman1990evidence}
VJ~Goldman, M~Santos, M~Shayegan, and JE~Cunningham.
\newblock Evidence for two-dimentional quantum wigner crystal.
\newblock {\em Physical review letters}, 65(17):2189, 1990.

\bibitem{jiang1990quantum}
HW~Jiang, RL~Willett, HL~Stormer, DC~Tsui, LN~Pfeiffer, and KW~West.
\newblock Quantum liquid versus electron solid around $\nu$= 1/5 landau-level filling.
\newblock {\em Physical review letters}, 65(5):633, 1990.

\bibitem{santos_observation_1992}
M.~B. Santos, Y.~W. Suen, M.~Shayegan, Y.~P. Li, L.~W. Engel, and D.~C. Tsui.
\newblock Observation of a reentrant insulating phase near the 1/3 fractional quantum {Hall} liquid in a two-dimensional hole system.
\newblock {\em Phys. Rev. Lett.}, 68(8):1188--1191, February 1992.
\newblock Publisher: American Physical Society.

\bibitem{grimes_evidence_1979}
C.~C. Grimes and G.~Adams.
\newblock Evidence for a {Liquid}-to-{Crystal} {Phase} {Transition} in a {Classical}, {Two}-{Dimensional} {Sheet} of {Electrons}.
\newblock {\em Phys. Rev. Lett.}, 42(12):795--798, March 1979.
\newblock Publisher: American Physical Society.

\bibitem{tsui_direct_2023}
Yen-Chen Tsui, Minhao He, Yuwen Hu, Ethan Lake, Taige Wang, Kenji Watanabe, Takashi Taniguchi, Michael~P. Zaletel, and Ali Yazdani.
\newblock Direct observation of a magnetic field-induced {Wigner} crystal, December 2023.
\newblock arXiv:2312.11632 [cond-mat].

\bibitem{xiang_quantum_2024}
Ziyu Xiang, Hongyuan Li, Jianghan Xiao, Mit~H. Naik, Zhehao Ge, Zehao He, Sudi Chen, Jiahui Nie, Shiyu Li, Yifan Jiang, Renee Sailus, Rounak Banerjee, Takashi Taniguchi, Kenji Watanabe, Sefaattin Tongay, Steven~G. Louie, Michael~F. Crommie, and Feng Wang.
\newblock Quantum {Melting} of a {Disordered} {Wigner} {Solid}, February 2024.
\newblock arXiv:2402.05456 [cond-mat].

\bibitem{kivelson_cooperative_1986}
Steven Kivelson, C.~Kallin, Daniel~P. Arovas, and J.~R. Schrieffer.
\newblock Cooperative ring exchange theory of the fractional quantized {Hall} effect.
\newblock {\em Phys. Rev. Lett.}, 56(8):873--876, February 1986.
\newblock Publisher: American Physical Society.

\bibitem{halperin_compatibility_1986}
B.~I. Halperin, Z.~Te\v{s}anovi\'c, and F.~Axel.
\newblock Compatibility of {Crystalline} {Order} and the {Quantized} {Hall} {Effect}.
\newblock {\em Phys. Rev. Lett.}, 57(7):922--922, August 1986.
\newblock Publisher: American Physical Society.

\bibitem{kivelson_cooperative_1987}
Steven Kivelson, C.~Kallin, Daniel~P. Arovas, and J.~Robert Schrieffer.
\newblock Cooperative ring exchange and the fractional quantum {Hall} effect.
\newblock {\em Phys. Rev. B}, 36(3):1620--1646, July 1987.
\newblock Publisher: American Physical Society.

\bibitem{tesanovic_hall_1989}
Zlatko Te{\v{s}}anovi{\'c}, Fran{\oldc{c}}oise Axel, and B.~I. Halperin.
\newblock ``{Hall} crystal'' versus {Wigner} crystal.
\newblock {\em Phys. Rev. B}, 39(12):8525--8551, April 1989.
\newblock Publisher: American Physical Society.

\bibitem{AHC_Yahui}
Boran Zhou, Hui Yang, and Ya-Hui Zhang.
\newblock Fractional quantum anomalous hall effect in rhombohedral multilayer graphene in the moir\'eless limit.
\newblock {\em Phys. Rev. Lett.}, 133:206504, Nov 2024.

\bibitem{AHC1}
Junkai Dong, Taige Wang, Tianle Wang, Tomohiro Soejima, Michael~P. Zaletel, Ashvin Vishwanath, and Daniel~E. Parker.
\newblock Anomalous hall crystals in rhombohedral multilayer graphene. i. interaction-driven chern bands and fractional quantum hall states at zero magnetic field.
\newblock {\em Phys. Rev. Lett.}, 133:206503, Nov 2024.

\bibitem{lu2024fractional}
Zhengguang Lu, Tonghang Han, Yuxuan Yao, Aidan~P Reddy, Jixiang Yang, Junseok Seo, Kenji Watanabe, Takashi Taniguchi, Liang Fu, and Long Ju.
\newblock Fractional quantum anomalous hall effect in multilayer graphene.
\newblock {\em Nature}, 626(8000):759--764, 2024.

\bibitem{widom1982thermodynamic}
A~Widom.
\newblock Thermodynamic derivation of the hall effect current.
\newblock {\em Physics Letters A}, 90(9):474, 1982.

\bibitem{streda1982theory}
P~Streda.
\newblock Theory of quantised hall conductivity in two dimensions.
\newblock {\em Journal of Physics C: Solid State Physics}, 15(22):L717, 1982.

\bibitem{kunz1986quantized}
H~Kunz.
\newblock Quantized currents and topological invariants for electrons in incommensurate potentials.
\newblock {\em Physical review letters}, 57(9):1095, 1986.

\bibitem{dirac1931quantised}
Paul Adrien~Maurice Dirac.
\newblock Quantised singularities in the electromagnetic field.
\newblock {\em Proceedings of the Royal Society of London. Series A, Containing Papers of a Mathematical and Physical Character}, 133(821):60--72, 1931.

\bibitem{Brauner_2024}
Tomá\v{s} Brauner.
\newblock {\em Effective Field Theory for Spontaneously Broken Symmetry}.
\newblock Springer International Publishing, 2024.

\bibitem{Note1}
As is the case for all pinning fields for symmetry broken states, the effect of this potential on the ground state is highly non-linear.

\bibitem{Note2}
As before, we take the thermodynamic limit before taking the $\lambda \to 0$ limit.

\bibitem{Note3}
{ When does a crystal follow the dragging potential? Let $\Delta $ be the gap induced by the dragging potential, $v$ the velocity of the crystal, and $a$ the lattice constant of the crystal. The adiabatic approximation holds when $\omega = v / a \ll \Delta /\hbar \DOTSB \protect \tmspace +\thickmuskip {.2777em}\DOTSB \protect \Relbar \protect \joinrel \Rightarrow \protect \tmspace +\thickmuskip {.2777em}v \ll a\Delta / \hbar $. For example, a Wigner crystal with a lattice constant of $10$ nm and a pinning gap of $1$ meV has $v = 1.5 \times 10^4$ m/s, showing that a very fast motion is possible. In the absence of an explicit dragging potential, the role of the dragging potential will be played by the self-consistent field generated by the crystal itself, and its strength will determine the regime of validity of our formula. }.

\bibitem{resta1998quantum}
Raffaele Resta.
\newblock Quantum-mechanical position operator in extended systems.
\newblock {\em Physical Review Letters}, 80(9):1800, 1998.

\bibitem{HARUKI2018}
Haruki Watanabe.
\newblock Insensitivity of bulk properties to the twisted boundary condition.
\newblock {\em Phys. Rev. B}, 98:155137, Oct 2018.

\bibitem{ManyBodyCN}
Qian Niu, D.~J. Thouless, and Yong-Shi Wu.
\newblock Quantized hall conductance as a topological invariant.
\newblock {\em Phys. Rev. B}, 31:3372--3377, Mar 1985.

\bibitem{Note4}
We use the convention such that for $eB>0$ the lowest Landau level has $C=1$.

\bibitem{Note5}
This argument was used to predict our result(Eq.~\ref {eq:bound_charge_formulation}) in Ref.~\cite {tesanovic_hall_1989}.

\bibitem{DanArovasQHENotes}
Daniel Arovas.
\newblock Lecture notes on quantum hall effect (a work in progress).

\bibitem{Note6}
We sincerely thank Patrick J. Ledwith for this simple but illuminating argument.

\bibitem{Note7}
In fact, our proof actually did not rely on the states having nonzero crystalline order parameter. Therefore, our derivation showed that fully filled Landau levels with $\nu \in \protect \mathbb {Z}$ carry zero current under moving pinning field, a trivial result given that the state is not affected by pinning.

\bibitem{HallCrystalTDHF}
Ganpathy Murthy.
\newblock Hall crystal states at $\mathit{\ensuremath{\nu}}\phantom{\rule{0ex}{0ex}}=\phantom{\rule{0ex}{0ex}}2$ and moderate landau level mixing.
\newblock {\em Phys. Rev. Lett.}, 85:1954--1957, Aug 2000.

\bibitem{Fukuyama_magnetic_1975}
H.~{Fukuyama}.
\newblock {Two-dimensional wigner crystal under magnetic field}.
\newblock {\em Solid State Communications}, 17(10):1323--1326, November 1975.

\bibitem{Bonsall-Maradudin}
Lynn Bonsall and A.~A. Maradudin.
\newblock Some static and dynamical properties of a two-dimensional wigner crystal.
\newblock {\em Phys. Rev. B}, 15:1959--1973, Feb 1977.

\bibitem{Note8}
This may differ from the cyclotron gap of the entire system, $\omega _c = eB/m_e$, where $m_e$ is the electron mass, as dictated by Kohn's theorem~\cite {kohn1961cyclotron}. This is because $\omega _c$ corresponds to the motion of all electrons at the same time, which differ from the motion of bound-charges considered here.

\bibitem{AHC4}
Junkai Dong, Ophelia~Evelyn Sommer, Tomohiro Soejima, Daniel~E. Parker, and Ashvin Vishwanath.
\newblock Phonons in electron crystals with berry curvature, 2025.

\bibitem{watanabe2020counting}
Haruki Watanabe.
\newblock Counting rules of nambu--goldstone modes.
\newblock {\em Annual Review of Condensed Matter Physics}, 11(1):169--187, 2020.

\bibitem{Note9}
The prescription for preparing this state was discussed in Ref.~\cite {Zeng_sliding}, including the proper method for gauge-fixing.

\bibitem{xiao2010berry}
Di~Xiao, Ming-Che Chang, and Qian Niu.
\newblock Berry phase effects on electronic properties.
\newblock {\em Reviews of modern physics}, 82(3):1959--2007, 2010.

\bibitem{chang1996berry}
Ming-Che Chang and Qian Niu.
\newblock Berry phase, hyperorbits, and the hofstadter spectrum: Semiclassical dynamics in magnetic bloch bands.
\newblock {\em Physical Review B}, 53(11):7010, 1996.

\bibitem{chang1995berry}
Ming-Che Chang and Qian Niu.
\newblock Berry phase, hyperorbits, and the hofstadter spectrum.
\newblock {\em Physical review letters}, 75(7):1348, 1995.

\bibitem{Note10}
The cyclotron frequency satisfies $\omega _c \ll 1/m\Omega ^\protect \mathrm {Pho}$ provided $\beta \Omega \ll 1$. Similar discussion on the role of $\Omega ^\protect \mathrm {Pho}$ can be found in Ref.~\cite {joy2025marchmeeting}.

\bibitem{Note11}
This breaking of continuous translation symmetries has been discussed in Refs.~\cite {fischler1979quark, metlitski2007schwinger, seiberg, seiberg2025anomalouscontinuoustranslations} as well as Appendix G of Ref.~\cite {geraedts2016half}.

\bibitem{Note12}
A similar anomaly-matching argument has been used to derive Luttinger's theorem for Fermi liquids~\cite {oshikawa2000topological, else2021non, cheng2023lieb}.

\bibitem{seiberg}
Nathan Seiberg.
\newblock {Ferromagnets, a new anomaly, instantons, and (noninvertible) continuous translations}.
\newblock {\em SciPost Phys.}, 18:063, 2025.

\bibitem{seiberg2025anomalouscontinuoustranslations}
Nathan Seiberg.
\newblock Anomalous continuous translations, 2025.

\bibitem{nagaosa2013topological}
Naoto Nagaosa and Yoshinori Tokura.
\newblock Topological properties and dynamics of magnetic skyrmions.
\newblock {\em Nature nanotechnology}, 8(12):899--911, 2013.

\bibitem{stone1996magnus}
Michael Stone.
\newblock Magnus force on skyrmions in ferromagnets and quantum hall systems.
\newblock {\em Physical Review B}, 53(24):16573, 1996.

\bibitem{gruner1988dynamics}
George Gr{\"u}ner.
\newblock The dynamics of charge-density waves.
\newblock {\em Reviews of modern physics}, 60(4):1129, 1988.

\bibitem{gruner2018density}
George Gruner.
\newblock {\em Density waves in solids}.
\newblock CRC press, 2018.

\bibitem{fisher1998collective}
Daniel~S Fisher.
\newblock Collective transport in random media: from superconductors to earthquakes.
\newblock {\em Physics reports}, 301(1-3):113--150, 1998.

\bibitem{brazovskii2004pinning}
Serguei Brazovskii and Thomas Nattermann.
\newblock Pinning and sliding of driven elastic systems: from domain walls to charge density waves.
\newblock {\em Advances in Physics}, 53(2):177--252, 2004.

\bibitem{reichhardt2016depinning}
Charles Reichhardt and CJ~Olson Reichhardt.
\newblock Depinning and nonequilibrium dynamic phases of particle assemblies driven over random and ordered substrates: a review.
\newblock {\em Reports on Progress in Physics}, 80(2):026501, 2016.

\bibitem{patri2024extended}
Adarsh~S Patri, Zhihuan Dong, and T~Senthil.
\newblock Extended quantum anomalous hall effect in moir{\'e} structures: Phase transitions and transport.
\newblock {\em Physical Review B}, 110(24):245115, 2024.

\bibitem{nandi2012exciton}
D~Nandi, ADK Finck, JP~Eisenstein, LN~Pfeiffer, and KW~West.
\newblock Exciton condensation and perfect coulomb drag.
\newblock {\em Nature}, 488(7412):481--484, 2012.

\bibitem{narozhny2016coulomb}
BN~Narozhny and A~Levchenko.
\newblock Coulomb drag.
\newblock {\em Reviews of Modern Physics}, 88(2):025003, 2016.

\bibitem{kohn1961cyclotron}
Walter Kohn.
\newblock Cyclotron resonance and de haas-van alphen oscillations of an interacting electron gas.
\newblock {\em Physical Review}, 123(4):1242, 1961.

\bibitem{Zeng_sliding}
Yongxin Zeng and Andrew~J. Millis.
\newblock Berry phase dynamics of sliding electron crystals, 2024.

\bibitem{joy2025marchmeeting}
Sandeep Joy and Brian Skinner.
\newblock Collective modes of a wigner crystal with berry curvature.
\newblock APS March meeting contributed talk, 2025.

\bibitem{fischler1979quark}
W~Fischler, J~Kogut, and Leonard Susskind.
\newblock Quark confinement in unusual environments.
\newblock {\em Physical Review D}, 19(4):1188, 1979.

\bibitem{metlitski2007schwinger}
Max~A Metlitski.
\newblock Is the schwinger model at finite density a crystal?
\newblock {\em Physical Review D—Particles, Fields, Gravitation, and Cosmology}, 75(4):045004, 2007.

\bibitem{geraedts2016half}
Scott~D Geraedts, Michael~P Zaletel, Roger~SK Mong, Max~A Metlitski, Ashvin Vishwanath, and Olexei~I Motrunich.
\newblock The half-filled landau level: The case for dirac composite fermions.
\newblock {\em Science}, 352(6282):197--201, 2016.

\bibitem{oshikawa2000topological}
Masaki Oshikawa.
\newblock Topological approach to luttinger's theorem and the fermi surface of a kondo lattice.
\newblock {\em Physical Review Letters}, 84(15):3370, 2000.

\bibitem{else2021non}
Dominic~V Else, Ryan Thorngren, and T~Senthil.
\newblock Non-fermi liquids as ersatz fermi liquids: General constraints on compressible metals.
\newblock {\em Physical Review X}, 11(2):021005, 2021.

\bibitem{cheng2023lieb}
Meng Cheng and Nathan Seiberg.
\newblock Lieb-schultz-mattis, luttinger, and't hooft-anomaly matching in lattice systems.
\newblock {\em SciPost Physics}, 15(2):051, 2023.

\bibitem{Note13}
We again use the convention that for $eB>0$ the Landau levels have $C=1$.

\bibitem{kohn1964theory}
Walter Kohn.
\newblock Theory of the insulating state.
\newblock {\em Physical review}, 133(1A):A171, 1964.

\bibitem{zang2011dynamics}
Jiadong Zang, Maxim Mostovoy, Jung~Hoon Han, and Naoto Nagaosa.
\newblock Dynamics of skyrmion crystals in metallic thin films.
\newblock {\em Physical review letters}, 107(13):136804, 2011.

\bibitem{watanabe2014noncommuting}
Haruki Watanabe and Hitoshi Murayama.
\newblock Noncommuting momenta of topological solitons.
\newblock {\em Physical Review Letters}, 112(19):191804, 2014.

\bibitem{Haldane1986}
F.D.M. Haldane.
\newblock Geometrical interpretation of momentum and crystal momentum of classical and quantum ferromagnetic heisenberg chains.
\newblock {\em Phys.\ Rev.\ Lett.}, 57:1488--1491, 1986.

\bibitem{Balakrishnan1985}
Radha Balakrishnan and A.R. Bishop.
\newblock Nonlinear excitations on a quantum ferromagnetic chain.
\newblock {\em Phys.\ Rev.\ Lett.}, 55:537--540, 1985.

\bibitem{Volovik1987}
G.E. Volovik.
\newblock Linear momentum in ferromagnets.
\newblock {\em J.\ Phys.\ C: Solid State Phys.}, 20:L83--L87, 1987.

\bibitem{Papanicolaou1991}
N.~Papanicolaou and T.N. Tomaras.
\newblock Dynamics of magnetic vortices.
\newblock {\em Nucl.\ Phys.\ B}, 360:425--462, 1991.

\bibitem{Floratos1992}
Floratos E.G.
\newblock A lagrangian resolution of the momentum problem for two-dimensional ferromagnets.
\newblock {\em Phys.\ Lett.\ B}, 279:117--123, 1992.

\bibitem{Banerjee1995}
R.~Banerjee and B.~Chakraborty.
\newblock Formulation of the landau--lifshitz model of ferromagnetism as a constrained dynamical system.
\newblock {\em Nucl.\ Phys.\ B}, 449:317--346, 1995.

\bibitem{Nair2004}
Nair V.P. and R.~Ray.
\newblock Some topological issues for ferromagnets and fluids.
\newblock {\em Nucl.\ Phys.\ B}, 676:659--675, 2004.

\bibitem{Tchernyshyov2015}
Oleg Tchernyshyov.
\newblock Conserved momenta of a ferromagnetic soliton.
\newblock {\em Ann.\ Phys.}, 363:98--113, 2015.

\bibitem{Dasgupta2018}
Sayak Dasgupta and Oleg Tchernyshyov.
\newblock Energy-momentum tensor of a ferromagnet.
\newblock {\em Phys.\ Rev.\ B}, 98:224401, 2018.

\bibitem{Di2021}
Xingjian Di and Oleg Tchernyshyov.
\newblock Conserved momenta of ferromagnetic solitons through the prism of differential geometry.
\newblock {\em SciPost Phys.}, 11:108, 2021.

\bibitem{Brauner2024}
Tom\'{a}\v{s} Brauner, Naoki Yamamoto, and Ryo Yokokura.
\newblock Dipole symmetries from the topology of the phase space and the constraints on the low-energy spectrum.
\newblock {\em SciPost Phys.}, 16(2):051, 2024.

\bibitem{Note14}
We thank Bert Halperin for walking us through this argument.

\end{thebibliography}

\let\temp\oldv
\let\oldv\v
\let\v\temp

\onecolumngrid
\appendix
\newpage

\section{Berry phase for the sliding crystal}
\label{app:berry_phase_for_crystals}

This Appendix shows that the Berry phase accumulated over a closed trajectory of states $\ket{\Psi(t)} = \ket{\v{u}(t);\v{\Phi}=0}$ (with $\v{u}(T) = \v{u}(0)$ and enclosed area $A_\circ$) is
\begin{equation}
    \label{eq:berryphase_app}
    \varphi_B = N_e eB(1-C\phi/\rho) A_{\circ},
\end{equation}
which is Eq.~\eqref{eq:berryphase} in the main text.
We compute the Berry curvature $\Omega(\v{u})$ as a Wilson loop
\begin{equation}
    \Omega(\v{u}) = \lim_{\delta u \to 0} \frac{\textrm{Im}\prod_{j=1}^{4}\braket{\v{u}^{(j)}|\v{u}^{(j+1)}}}{\delta u^2}
    \label{eq:BerryCurvatureDef}
\end{equation}
in which we take $\v{u}^{(1)} = \v{u}^{(5)} = \v{u}, \v{u}^{(2)} =\v{u}+(\delta u,0), \v{u}^{(3)} = \v{u}+(\delta u,\delta u), \v{u}^{(4)} =\v{u}+(0,\delta u)$, with the flux fixed at $\v{\Phi}=0$ throughout.

Using Eq.~\eqref{eq:translate_to_zero_COM},
\begin{equation}
    \ket{\v{u};\v{\Phi}=0} = \hat{T}_{\v u} \ket{\v u=0; \v\Phi=-eB(L_xu_y,-L_yu_x)},
\end{equation}
the overlaps become
\begin{equation}
   \prod_{j=1}^4 \bra{0; \v{\Phi}^{(j)}}
    \hat{T}^\dagger_{\v{u}^{(j)}}\hat{T}_{\v{u}^{(j+1)}} \ket{0; \v{\Phi}^{(j+1)}}
    = e^{iN_e eB\delta u^2}\prod_{j=1}^4 \bra{0; \v{\Phi}^{(j)}}
    \hat{T}_{\v{u}^{(j+1)}-\v{u^{(j)}}} \ket{0; \v{\Phi}^{(j+1)}},
    \label{eq:pancharatnam}
\end{equation}
where we have used $\hat{T}_{\v{u}_1}\hat{T}_{\v{u}_2} = e^{i eB \v{u}_1 \wedge \v{u}_2} \hat{T}_{\v{u}_2} \hat{T}_{\v{u}_1} = e^{i \frac{eB}{2} \v{u}_1 \wedge \v{u}_2} T_{\v{u}_1+\v{u}_2}$, the identity $\prod_{j=1}^4 e^{i N_e eB \v{u}_j \wedge \v{u}_{j+1}/2} = e^{iN_e eB\delta u^2}$, and defined $\v{\Phi}^{(j)} = -eB(L_xu_y^{(j)},-L_yu_x^{(j)})$. 

The exponential prefactor in Eq.~\eqref{eq:pancharatnam} corresponds to the standard Aharonov-Bohm phase accumulated for a particle in a magnetic field, while the product will give a phase proportional to the many-body Chern number.
To see this, we will examine one of the overlaps explicitly; the others are directly analogous. Consider
\begin{equation}
    O_{12} = \bra{0; \v{\Phi}^{(1)}}
    \hat{T}_{\v{u}^{(2)}-\v{u^{(1)}}} \ket{0; \v{\Phi}^{(2)}} = \bra{0; -eB(L_xu_y,-L_yu_x)}
    \hat{T}_{(\delta u, 0)} \ket{0; -eB(L_xu_y,-L_y(u_x+\delta u))}.
\end{equation}
We now define
\begin{equation}
    \ket{U_{\v\Phi}} = e^{-i \sum_{a}( \Phi_x\cdot \hat{x}_{a}/L_x + \Phi_y\cdot \hat{y}_{a}/L_y)}\ket{0; \v{\Phi}},
\end{equation}
which are the many-body Bloch states that are periodic under magnetic translations by $\v{L}_{1,2}$, i.e. $\ket{U_{\v{\Phi}}} \in \mathcal{H}_{\v{\Phi} =\v{0}}.$ Here $a \in [1,N_e]$ labels the different particles. In terms of the Bloch states, 
\begin{equation}
\begin{aligned}
   O_{12} = \bra{0; \v{\Phi}^{(1)}}
    \hat{T}_{\v{u}^{(2)}-\v{u^{(1)}}} \ket{0; \v{\Phi}^{(2)}}
    &= e^{i N_e \Phi^{(2)}_x\delta u/L_x}\braket{U_{\v\Phi^{(1)}}|e^{i eB \delta u\sum_a \hat{y}_a} \hat{T}_{(\delta u, 0)} |U_{\v\Phi^{(2)}}}.
\end{aligned}
\label{eq:overlap}
\end{equation}
The Berry connection is determined by the linear order term in $\delta u$. We first expand the operator in the middle:
\begin{equation}
    e^{i eB \delta u\sum_a y_a} \hat{T}_{(\delta u, 0)} = 1 + i \delta u\left(eB \sum_a \hat{y}_a + \hat{R}_{x,a} \right)  + O(\delta u^2) = 1 + i \delta u \sum_a \hat{\pi}_{x,a} + O(\delta u^2).
\end{equation}
Here we have used that the guiding center for particle $a$, 
$\v R_a = \v p_a-e\v A_a+ e B(\hat{z}\times \v r_a)$, is the generator of magnetic translations, and the definition of the kinematic momentum $\v\pi_a = \v p_a -e \v A_a$. This can be understood physically as follows: since both $\ket{U_{\v \Phi}}'s$ are periodic under magnetic translations, the operator in the middle must be a translation operator that commutes with magnetic translations, which can only be generated by kinematic momentum as $[\v{R_a}, \v{\pi}_b] = 0$.

Now we compute the entire overlap $O_{12}$ to order $\delta u$:
\begin{equation}
\begin{aligned}
    \bra{0; \v{\Phi}^{(1)}}
    \hat{T}_{\v{u}^{(2)}-\v{u^{(1)}}} \ket{0; \v{\Phi}^{(2)}} &= 1 + i \left(\frac{N_e \Phi_x^{(2)}}{L_x} -i \braket{\partial_{\Phi^{(2)}_y}U_{\v{\Phi}^{(2)}}|U_{\v{\Phi}^{(2)}}} \frac{\partial \Phi^{(2)}_y}{\partial \delta u} + \braket{U_{\v\Phi^{(2)}}|\sum_a \pi_a|U_{\v\Phi^{(2)}}}\right)\delta u  + O(\delta u^2)\\
    & = 1 + i \left( -ieB L_y\braket{\partial_{\Phi^{(2)}_y}U_{\v{\Phi}^{(2)}}|U_{\v{\Phi}^{(2)}}} + \braket{0;\v{\Phi}^{(2)}|\sum_a \pi_a|0;\v{\Phi}^{(2)}}\right)\delta u  + O(\delta u^2)\\
    & = 1 + i eB L_y \tilde{A}_{y}(\v{\Phi}^{(2)})\delta u  + O(\delta u^2).
\end{aligned}
\label{eq:overlapsimplification}
\end{equation}
Here in the second line we collected the first and third term using the definition of the many-body Bloch state. In the third line we simplified the first term using the many-body Berry connection in the flux torus $\tilde{A}_a(\v\Phi) = -i\braket{U_{\v \Phi}|\partial_{\Phi_a} U_{\v \Phi}}$, and the expectation value of the kinetic momentum $\braket{0;\v{\Phi}^{(2)}|\sum_a \pi_a|0;\v{\Phi}^{(2)}}=0$. This expectation vanishes since it can be written as a derivative of a local gauge-invariant operator $\pi^2/2$ with respect to the flux $(\Phi_x, \Phi_y)$~\cite{HARUKI2018}.
Physically it makes sense for the expectation to vanish: in the case of a Galilean symmetric system one can identify the mechanical momentum with the center-of-mass velocity. Given that we have pinned the crystal, it should not have the freedom to slide.

Similar to the calculation above, we have
\begin{equation}
    O_{23} = \bra{0; \v{\Phi}^{(2)}}
    \hat{T}_{\v{u}^{(3)}-\v{u^{(2)}}} \ket{0; \v{\Phi}^{(3)}}
    = 1 + i eB L_x \tilde{A}_{x}(\v{\Phi}^{(3)})\delta u  + O(\delta u^2).
\end{equation}
The rest of the overlaps can be extracted by the same method.

We now calculate the product in Eq.~(\ref{eq:pancharatnam}). Using Green's theorem, we obtain
\begin{equation}
\begin{aligned}
    \prod_{j=1}^4 \bra{0; \v{\Phi}^{(j)}}
    \hat{T}_{\v{u}^{(j+1)}-\v{u^{(j)}}} \ket{0; \v{\Phi}^{(j+1)}} 
    =O_{12}O_{23}O_{31}O_{41}
    = \exp\left(-i e B (L_y\partial_{u_y} \tilde{A}_y(\v{\Phi}) +L_x\partial_{u_x} \tilde{A}_x(\v{\Phi}) )\delta u^2 + O(\delta u^3) \right).
\end{aligned}
\end{equation}
Recalling that $\v\Phi = -eB(L_xu_y,-L_yu_x)$, we can thus simplify
\begin{align}
    \label{eq:final_simplification}
    \prod_{j=1}^4 \bra{0; \v{\Phi}^{(j)}}
    \hat{T}_{\v{u}^{(j+1)}-\v{u^{(j)}}} \ket{0; \v{\Phi}^{(j+1)}}
    &= \exp\left(-ie^2B^2L_xL_y(-\partial_{\Phi_x}\tilde{A}_{y}(\v{\Phi}) + \partial_{\Phi_y}\tilde{A}_{x}(\v{\Phi}))\delta u^2
    + O(\delta u^3) \right)\\ 
    &= \exp\left(-ie^2B^2L_xL_y \tilde{\Omega}(\v{\Phi})\delta u^2 + O(\delta u^3)\right),
\end{align}
where $\tilde{\Omega}(\v{\Phi}) = -\partial_{\Phi_x}\tilde{A}_{y}(\v{\Phi}) + \partial_{\Phi_y}\tilde{A}_{x}(\v{\Phi})$ is the many-body Berry curvature on the flux torus at constant $\v{u}$~\footnote{We again use the convention that for $eB>0$ the Landau levels have $C=1$.}. 
Collecting Eq.~(\ref{eq:pancharatnam}), Eq.~(\ref{eq:final_simplification}) and substituting into Eq.~(\ref{eq:BerryCurvatureDef}), we obtain
\begin{equation}
    \Omega(\v{u})= \lim_{\delta u \to 0} \frac{1}{\delta u^2}
    \exp\left(iN_e eB\delta u^2-ie^2B^2L_xL_y \tilde{\Omega}(\v{\Phi})\delta u^2 + O(\delta u^3)\right)
    = eB \left[ N_e - 2\pi \tilde{\Omega}(\v{\Phi}) N_{\Phi} \right],
\end{equation}
with $N_\Phi = eBL_xL_y/2\pi$. Finally, we use the fact that the many-body Berry curvature on the flux torus is essentially constant, $2\pi \tilde{\Omega}(\v{\Phi}) = C + O(e^{-L})$\cite{HARUKI2018}, to find
\begin{equation}
    \Omega(\v{u}) = eB(N_e - N_\phi C) = N_e eB(1-C\phi/\rho).
\end{equation}
Eq.~(\ref{eq:berryphase_app}) then naturally follows.

\section{Response of Hall crystals}
\label{app:hall_crystal_response}

In this section, we compute the response of (pinned) Hall crystals to electromagnetic fields. Throughout, we only consider the coupling between crystalline degrees of freedom and electromagnetic fields. Electrons pinned by the magnetic field (e.g. filled Landau levels) can contribute to the response at finite frequencies, but we ignore their effects. Our findings can be summarized as follows:
\begin{enumerate}
    \item Pinned Hall crystals have nonzero Hall response but zero longtudinal response i.e. it is an electric insulator.
    \item Unpinned Hall crystal has Hall conductivity consistent with Galilean invariance.
    \item Anomalous Hall crystals pick up a correction due to momentum-space Berry phase and velocity-dependent polarization.
    \item Full Hall crystals pick up a correction due to velocity-dependent polarization.
\end{enumerate}

Consider the low frequency $\v k=\v 0$ electromagnetic response of a Hall crystal with phonons, which may described by the following action 
\begin{equation}
\label{eq:phonon_action_app}
    S^\mathrm{phonon}[\v u]=\frac{\rho}{2}\int d td^2\v r \left(~-\Lambda\v u^2+\beta\dot{\v u}\times \v u+m\dot{\v u}^2+\Omega^{\mathrm{pho}} m^2\ddot{\v u}\times \dot{\v u}\right)=\frac{\rho L_xL_y}{2}\int\frac{d \omega}{2\pi}~u^a(\omega) u^b(-\omega) D_{ab}^{-1}(\omega).
\end{equation}
Here and below we use Einstein notation with spatial indices $a,b,c\dots$, $\Lambda$ is the strength of a pinning field, $\beta=s eB/\rho A_\mathrm{uc}$ is an effective Lorentz force, $m$ is the effective mass, and $\Omega^{\mathrm{pho}}$ is the phonon Berry curvature. As in the main text, 
$s/A_\mathrm{uc}=\rho-C\phi=\rho \tilde{s}$ where $s$ is the bound charge topological invariant of the Widom-St\oldv{r}eda formula, and $\phi=N_\Phi/L_xL_y$ is the flux density. The electromagnetic coupling between the vector potential $\v A(\omega)=\frac{\v E(\omega)}{i\omega}$ (working in temporal gauge) and the polarization  takes the form
\begin{equation}
    S^\mathrm{coupling}[\v u,\v A]=\rho \tilde{s} e\int d t d^2\v r \; \v u\cdot\v E
    + \rho \alpha e \int dt d^2 \v{r} \;  \v{\dot{u}}\times \v{E},
\end{equation}
since the change of the polarization is given by $\tilde{s}\rho e$. The second term arises from the velocity dependence of the polarization. For example, this can be caused by the Lorentz force acting on excitons.
Including the Chern-Simons term for the many-body Chern number $C$, the leading order action is therefore 
\begin{equation}
    S[\v u,\v A]=S^\mathrm{phonon}[\v u]+S^\mathrm{coupling}[\v u,\v A]+\frac{e^2 C}{4\pi}\int d t d^2 \v r\; \v A\times \dot{\v A},
\end{equation}
which captures the low momentum/frequency response where we neglect the electromagnetic response of the non-phonon degrees of freedom beyond the Chern-Simons term, $\Delta S^\mathrm{EM}_\mathrm{background}$, which dominantly contributes via the subleading background permittivity term $\epsilon_\mathrm{background}^{ab}(\omega=0)\neq0$. We also neglect the nonlinearities and higher order frequency effects from the phonon and phonon-EM coupling actions.
The coefficient of the Chern-Simons term ensures quantized response for pinned crystals.
To study linear response to leading order in $\omega$, it is sufficient to keep quadratic order terms in $\v u,\v A$ and lowest order in $\omega$. The electromagnetic response is straightforward to obtain by integrating out $\v u$: 
\begin{align}
    &\frac{\rho L_xL_y}{2}\int\frac{d \omega}{2\pi}~u^a(\omega) u^b(-\omega) D_{ab}^{-1}(\omega)+i\omega \tilde{s}  e  [u^a(-\omega)A_a(\omega)-u^a(\omega)A_a(-\omega)]
    +\omega^2 \alpha e \epsilon_{ab}[u_a(-\omega)A_b(\omega)+u_a(\omega)A_b(-\omega)]
    \\ 
    &=\frac{\rho L_xL_y}{2}\int\frac{d \omega}{2\pi}~\Bigg\{[u^a(\omega)
    +(i\omega \tilde{s} eA_c(\omega) + \omega^2 \alpha  e A_e(\omega) \epsilon^{ce}
    )
    D^{ca}(\omega)]\\
    &\hspace{1in}\times[u^b(-\omega)
    +(-i\omega \tilde{s} eA_d(-\omega) + \omega^2 \alpha e A_f(-\omega) \epsilon^{df}
    )
    D^{bd}(\omega)] D_{ab}^{-1}(\omega)
    \\
    &\hspace{1in}-e^2\left(
    \tilde{s}^2e \omega^2 D^{ab}(\omega)
    +i2\omega^3\tilde{s} \alpha D^{ac}(\omega) \epsilon^{cb}
    +\omega^4 \alpha^2 D^{cd}(\omega) \epsilon^{ac}\epsilon^{db}
    \right)A_a(\omega)A_b(-\omega)
    \Bigg\}.
\end{align}
For quadratic actions there is no need to distinguish the effective and bare actions, and whether by solving the Euler-Lagrange equations, or via completing the square and shifting path integral integration variable, the effective electromagnetic action takes the form
\begin{equation}
    S^\mathrm{EM}=\frac{e^2L_xL_y}{2}\int\frac{d \omega}{2\pi}  \left[-\rho \tilde{s}^2\omega^2D^{ab}(\omega)
    +i2\omega^3\rho\tilde{s}\alpha D^{ac}(\omega) \epsilon^{cb}
    +\omega^4 \rho\alpha^2 D^{cd} \epsilon^{ac}\epsilon^{db}
    +\frac{i \omega C}{2\pi} \epsilon^{ab}\right]A_a(\omega)A_b(-\omega).
\end{equation}
This produces a current
\begin{equation}
j^a=\frac{1}{L_xL_y}\frac{\delta S^\mathrm{EM}}{\delta A_a(-\omega)}=e^2\left(
i\rho \tilde{s}^2\omega D^{ab}(\omega)
    +2\omega^2\rho\tilde{s}\alpha D^{ac}(\omega) \epsilon^{cb}
    -i\omega^3 \rho\alpha^2 D^{cd} \epsilon^{ac}\epsilon^{db}
+\frac{ C\epsilon^{ab}}{2\pi}\right) E_b(\omega)
\end{equation}
which implies the conductivity tensor is
\begin{equation}
    \sigma^{ab}(\omega)=e^2\left(
i\rho \tilde{s}^2\omega D^{ab}(\omega)
    +2\omega^2\rho\tilde{s}\alpha D^{ac}(\omega) \epsilon^{cb}
    -i\omega^3 \rho\alpha^2 D^{cd} \epsilon^{ac}\epsilon^{db}
+\frac{ C\epsilon^{ab}}{2\pi}\right)
\end{equation}
From Eq.~\eqref{eq:phonon_action_app}, the inverse phonon Green's function is
\begin{align}
    D_{ab}^{-1}(\omega)&=(-\Lambda+m\omega^2)\delta_{ab}-i\omega\epsilon_{ab}(\beta+m^2\omega^2\Omega^\mathrm{pho})\\ 
     D^{ab}(\omega)&=\frac{1}{(-\Lambda+m\omega^2)^2-\omega^2(\beta+m^2\omega^2\Omega^\mathrm{pho})^2}[(-\Lambda+m\omega^2)\delta_{ab}+i\omega\epsilon_{ab}(\beta+m^2\omega^2\Omega^\mathrm{pho})].
\end{align}
Combining these results we get
\begin{align}
        \sigma^{ab}(\omega)=
        \frac{e^2\rho}{(-\Lambda+m\omega^2)^2-\omega^2(\beta+m^2\omega^2\Omega^\mathrm{pho})^2}
(
&[(-\Lambda+m\omega^2)(i \tilde{s}^2 \omega  - i\omega^3  \alpha^2) + i2\omega^3  \tilde{s}\alpha(\beta + m^2\omega^2\Omega^\mathrm{pho})
]\delta_{ab}
\\
+&
[(\beta+m^2\omega^2\Omega^\mathrm{pho})(- \tilde{s}^2 \omega^2 + \omega^4 \alpha^2) - 2 \omega^2  \tilde{s}\alpha(-\Lambda + m\omega^2)]\epsilon^{ab}
)\\
+&\frac{e^2C\epsilon^{ab}}{2\pi}.
\end{align}
To understand this response, we consider four physical cases.

\begin{enumerate}
    \item If $\Lambda$ is finite, as $\omega\to 0$ the phonons do not contribute to the conductivity. As expected we find an insulator with
\begin{equation}
    \sigma^{ab}(\omega\approx 0)=
    \frac{e^2\rho}{\Lambda^2}\left(-i\Lambda \tilde{s}^2 \omega \delta^{ab}
    -
    (\beta  \tilde{s}^2 - 2 \tilde{s}{\alpha \Lambda})\omega^2  \epsilon^{ab}\right)
    +
    \frac{e^2C}{2\pi}\epsilon^{ab} \to \frac{e^2C}{2\pi}\epsilon^{ab},
\end{equation}
where we kept terms up to quadratic order in $\omega$.
Notably, when $\tilde{s} =0$, the conductivity vanishes up to linear order. This can be used as a signature of full Hall crystals.

\item If $\Lambda=0$, but $\beta\neq0$, the conductivity becomes 
\begin{equation}
    \sigma^{ab}(\omega\to 0)=\frac{e^2\epsilon^{ab}}{2\pi}\left(C+\frac{2\pi\rho\tilde{s}^2}{\beta}\right)=\frac{e^2\epsilon^{ab}}{2\pi}\left(C+(\rho-C\phi)\frac{2\pi}{eB}\right)=\frac{e\rho}{B}\epsilon^{ab}.
\end{equation}
Thus in the unpinned case with a magnetic field, the Hall conductivity is indistinguishable from a free electron gas, reflecting the emergent Galilean symmetry of the action.

\item If $\Lambda=\beta=0$ but $\tilde{s}=1$, which describes an (anomalous Hall) crystal without magnetic field, the conductivity takes the form 
\begin{equation}
\label{eq:AHC_cond}
    \sigma^{ab}(\omega\to0)=\frac{i\rho e^2 }{m (\omega+i0^+)} \delta^{ab}+\frac{e^2}{2\pi}\left(C-2\pi\rho \Omega^\mathrm{pho} - 4\pi \rho \frac{\alpha}{m}\right)\epsilon^{ab}.
\end{equation}
The longitudinal conductivity of crystal diverges, giving rise to the Drude weight. The Hall conductivity has two contributions. The first contribution $C-2\pi\rho \Omega^\mathrm{pho}$ has already been discussed in Ref.~\cite{Zeng_sliding}, and it indicates that the Hall coefficient is modified from the many body Chern number by the presence of center of mass momentum Berry curvature $\Omega^\mathrm{pho}$, which is not expected to be quantized.
The second term, on the other hand, arises from the coupling of velocity with polarization.
\item Finally, when $\Lambda = s = 0$, we have
\begin{equation}
    \sigma^{ab}(\omega \approx 0) = \frac{e^2C}{2\pi}\epsilon^{ab} + e^2 \alpha^2 \rho
    \left[
    \frac{-i\omega}{m}{\delta^{ab} + \Omega^\mathrm{pho} \omega^2 \epsilon^{ab}}
    \right].
\end{equation}
If we reintroduce the next-leading order term in the EM action (from the permittivity of the non-phonon background degrees-of freedom)
\begin{equation}
\Delta S^\mathrm{EM}_\mathrm{background}= \int d t d\v r\;   \frac{1}{2}\epsilon_\mathrm{background}^{ab}\dot{ A}_a\dot{A}_b,
\end{equation}
then the current is modified by
$\Delta j^a(\omega\simeq0)=\frac{1}{L_xL_y}\frac{\delta  \Delta S^\mathrm{EM}_\mathrm{background}}{\delta A_a(-\omega)}=\epsilon_{\mathrm{bg}}^{ab}\omega^2 A_b(\omega)=-i\omega \epsilon_\mathrm{bg}^{ab}E_b (\omega)$. Hence the $\sigma^{ab}(\omega\simeq0)\propto \omega$ response when $s=\Lambda=0$ receives contributions beyond the phonon degrees of freedom.
\end{enumerate}

\section{Fractional Hall crystals}
\label{app:fractional_hall_crystals}

This section extends the results in the main text to the case of fractional Hall crystals. The first subsection shows Eq.~\eqref{eq:crystalline_current_formula} also holds for fractionalized crystalline systems. The argument is the same as the main text, but requires a small amount of additional bookkeeping. The second subsection discusses how emanant discrete translation symmetry is modified in the fractionalized case.

\subsection{Fractionalized Crystalline Current}
A fractional Hall crystal is a crystal with fractional Hall conductivity~\cite{tesanovic_hall_1989}. Such phase of matter has a topological order, which implies topological degeneracy of the ground states even in the presence of the pinning potential. We must modify our analysis to account for topological degeneracy of the ground states. 

Consider a fractional crystalline insulator whose Hamiltonian in the presence of a pinning field and background flux is
\begin{equation}
    \hat{H}(\v{u}, \v{\Phi}) = \sum_j \hat{h}(\v{p}_j - e \v{A}) + \sum_{i < j} \hat{V}(\v{r}_i - \v{r}_j)
    + \sum_j \lambda U(\v{r}_j - \v{u})
    \text{     where   } \v{A} = (-By, 0) + \left(\frac{\Phi_x}{eL_x}, \frac{\Phi_y}{eL_y}\right).
\end{equation}

To define its many-body Chern number, we consider an adiabatic flux threading process 
starting from an initial state $\ket{\Psi_0} \in \mathcal{H}_{\v{\Phi}=0}$ and time-dependent flux $\v{\Phi}(t), \v{\Phi}(0) = 0$ according to the Hamiltonian $H(0, \v{\Phi}(t))$. Concretely, denote the solution to this process by $\ket{\Psi_t} \in \mathcal{H}_{\v{\Phi}=0}$, we have
\begin{equation}
    \hat{H}(0, \v{\Phi}(t)) \ket{\Psi_t} = i\hbar \frac{\partial}{\partial t} \ket{\Psi_t}.
\end{equation}

This convention fixes the boundary condition constant while changing the gauge field to attain different flux sectors.
To go back to the convention in the main text, where different flux sectors corresponded to different boundary conditions on the torus, we consider the boundary condition changing operator~\cite{kohn1964theory} $\hat{U}[\v{\Phi}(t)]: \mathcal{H}_{\v{0}} \to \mathcal{H}_{\v{\Phi}(t)}$:
\begin{equation}
\hat{U}[\v \Phi(t)] = e^{-i \sum_{a=1}^{N_e} \sum_{\mu} \Phi_{a,\mu}(t) \hat{r}_{a,\mu}/L_\mu},
\end{equation}
where $\hat{\v{r}}_{a}$ is the position of the $a$\textsuperscript{th} electron and $\mu = x,y$. It satisfies
\begin{equation}
    \hat{U}[\v{\Phi}(t)]^\dagger \hat{H}(\v{u}, \v{\Phi}(t))     \hat{U}[\v{\Phi}(t)] = \hat{H}(\v{u}, \v{0}).
\end{equation}
We can thus find the state with the correct boundary condition to be
\begin{equation}
    \ket{\Psi_t; \v{\Phi}(t)} \equiv \hat{U}[\v{\Phi}(t)] \ket{\Psi_t} \in \mathcal{H}_{\v{\Phi}(t)},
\end{equation}
where we denote the boundary condition of the state explicitly. We will appreviate the second argument when there is no confusion.

Denoting the solution to this adiabatic process by $\ket{\Psi_0, \v{\Phi}(t)}$, the many-body Chern number is
\begin{equation}
    C = 2\pi \frac{d P_x}{dt} \left(\frac{d\Phi_y}{dt}\right)^{-1}
    \text{      where }
    P_x = \frac{L_x}{2\pi} \mathrm{Im} \log \bra{\Psi_t, \v{\Phi}(t)} e^{2\pi i \hat{x} / L_x}\ket{\Psi_t, \v{\Phi}(t)}.
    \label{eq:fractional_many_body_chern}
\end{equation}
This is directly analogous to Eq.~\eqref{eq:many_body_chern_number_from_polarization}, except we were careful to define it for an adiabatic process due to the degeneracy of the ground state subspace.

We can analogously define a sliding fractional Hall crystal as the solution to time-dependent Sch\"odinger equation for $\v{u} = \v{u}(t), \v{\Phi} = \v{0}$, starting from the same initial state $\ket{\Psi_0} \in \v{H}_{\v{0}}$. We denote the solution as $\ket{\Psi_t;\v{u} = \v{u}(t)} \in \mathcal{H}_{\v{0}}$. Now, we note that the magnetic translation operator $\hat{T}_{\v{u}}$ acts as
\begin{equation}
   \hat{T}_{\v{u}} H(\v{u}, 0)) \hat{T}_{\v{u}}^{-1} = 
   H(0, 0),
\end{equation}
while changing the boundary condition from $\mathcal{H}_{\v{0}}$ to  $\mathcal{H}_{\v{\Phi}}, \v{\Phi} = -eB(L_x u_y(t), - L_y u_x(t))$.
From this, we see that
\begin{equation}
    \ket{\Psi_t;\v{u} = \v{u}(t)} = \hat{T}_{\v{u}(t)} \ket{\Psi_t;\v{\Phi} = \v\Phi(t)}.
\end{equation}
As in the derivation in the main text, the change in the many-body polarization is now split into the part from the magnetic translation operator and the part from flux threading. The rest of the derivation in the main text goes through, replacing the definition of the many-body Chern number by Eq.~\ref{eq:fractional_many_body_chern}.

We can similarly compute the Berry phase as
\begin{equation}
    \prod \braket{\Psi_0; \v{u}(t_{j})|{\Psi_0; \v{u}(t_{j+1})}} = 
    \prod_j \braket{\Psi_0; \v{\Phi}(t_{j})|T^\dagger_{\v{u}_{j}}T_{\v{u}_{j+1}}|{\Psi_0; \v{\Phi}(t_{j+1})}} = e^{iN_e B \delta u^2}\prod_j \braket{\Psi_0; \v{\Phi}(t_{j})|T_{\v{u}_{j+1} - \v{u}_{j}}|{\Psi_0; \v{\Phi}(t_{j+1})}}.
\end{equation}
which gives us a contribution arising from many-body Berry curvature $\tilde{\Omega} = C/2\pi$, where $C$ may now be fractional.

\subsection{Implication for translation algebra}
An interesting consequence of the degeneracy can be seen in the algebra of translations. Recall that in the case of a unique ground state, we had $\ket{\v{u};\v{\Phi}} = \hat{T}_{\v{u}}\ket{0; \v{\Phi}+ \Delta \v{\Phi}}$, which implies
\begin{equation}
    \ket{\v{u}+\v{b}_j;\v{\Phi}} \propto \hat{T}_{\v{b}_j}\ket{\v{u}; \v{\Phi}},
\end{equation}
for the minimal translations $\v{b}_j$. Using this, we can compute the Berry phase around a rectangle generated by $\v{b}_1, \v{b}_2$ as
\begin{align}
    &\braket{\v{u};\v{\Phi}| \v{u} + \v{b}_1;\v{\Phi}}
    \braket{\v{u} + \v{b}_1;\v{\Phi}| \v{u} + \v{b}_1 + \v{b}_2;\v{\Phi}}
    \braket{\v{u} + \v{b}_1 + \v{b}_2;\v{\Phi}| \v{u} + \v{b}_2;\v{\Phi}}
    \braket{\v{u} + \v{b}_2;\v{\Phi}| \v{u};\v{\Phi}}
    \\
    =& e^{iN_e B \v{b}_1 \wedge \v{b}_2}
    |\braket{\v{u}; \v{\Phi}| \v{u} ;\v{\Phi}}|^4
    = e^{iN_e B \v{b}_1 \wedge \v{b}_2} = e^{i 2\pi\frac{N_e}{N_\Phi}} = e^{i2\pi\nu}.
\end{align}
In fact, this is simply a rephrasing of the anomaly matching constraint Eq.~\eqref{eq:anomaly_matching_constraint}.
On the other hand, for a fractional state we saw the Berry phase for the minimal loop should be given by
\begin{equation}
     \braket{\Psi_0; \v{u}| \Psi_0; \v{u}+\v{b}_1}
      \braket{\Psi_0; \v{u} + \v{b}_1| \Psi_0; \v{u}+\v{b}_1 + \v{b}_2}
       \braket{\Psi_0; \v{u}+\v{b}_1 + \v{b}_2| \Psi_0; \v{u}+\v{b}_2}
    \braket{\Psi_0; \v{u} + \v{b}_2| \Psi_0; \v{u}} =  e^{i2\pi (\nu - C)}.
\end{equation}
For fractional $C$, these do not match. The resolution comes from the fact
\begin{equation}
    \ket{\v{u}+\v{b}_j;\v{\Phi}} \neq  \hat{T}_{\v{b}_j}\ket{\v{u}; \v{\Phi}},
    \label{eq:fractional_states_do_not_translate_well}
\end{equation}
at least for one of $\v{b}_1$ or $\v{b}_2$.
This is a natural consequence of the fractional nature of the state, which requires $\ket{\Psi_0;\v{\Phi}} \neq \ket{\Psi_0;\v{\Phi} + 2\pi \v{e}_j}$, as $\v{\Phi}$ is changed adiabatically.

Eq.~\eqref{eq:fractional_states_do_not_translate_well} implies translations by $\hat{T}_{\v{b}_j}$ do not correspond to the translation of a sliding crystal. In the language of emanant symmetries, this means that microscopic translation operators do not get embedded as translation operators in the IR. Instead, it is embedded as an operator $\hat{U}_i$ that acts nontrivially on the topological sector:
\begin{equation}
    \hat{T}_{\v{b}_i}^\mathrm{UV} \to \hat{T}^\mathrm{IR}_{\v{b}_i} \hat{U}_i.
\end{equation}
For the $\nu = 1/3$ Laughlin state, for example, we can choose a basis on the topological sectosr such that
\begin{equation}
    \hat{U}_1 = \begin{pmatrix}
        1 & 0 & 0 \\
        0 & e^{i\frac{2\pi}{3}} & 0 \\
        0 & 0 & e^{-i\frac{2\pi}{3}}
    \end{pmatrix},
    \quad
    \hat{U}_2 = \begin{pmatrix}
        0 & 1 & 0 \\
        0 & 0 & 1 \\
        1 & 0 & 0
    \end{pmatrix}.
\end{equation}
Then, UV translation operators that map to topological sector preserving IR operators are given by
\begin{equation}
    \hat T_{\v{b}_1}^\mathrm{UV} \to \hat T_{\v{b}_1}^\mathrm{IR}, \quad
    (\hat T^\mathrm{UV}_{\v{b}_2})^3 \to \hat T^{\mathrm{IR}}_{3\v{b}_2}.
\end{equation}
This changes the anomaly matching condition to
\begin{equation}
    \beta = B(\nu - m) \phi / \rho \text{   where } m \in \frac{\mathbb{Z}}{3},
\end{equation}
which is compatible with the fractionalization of the Chern number $C=1/3$.

For interested readers, we note that a similar analysis  on the role of discrete translation symmetry was performed for the Dirac-composite Fermi liquid in the Appendix G of Ref.~\cite{geraedts2016half}.

\section{Emanant symmetry of skyrmions}
\label{app:skrymion}
In this section we discuss and contrast phonons in skyrmion crystals~\cite{nagaosa2013topological, zang2011dynamics} with the phonons in full Hall crystals in the main text. For skyrmions, it has been found that, although there are commuting lattice translations, the low energy action has a Berry phase term. Thus it appears as if the translations are non-commutative~\cite{stone1996magnus, watanabe2014noncommuting}. In the main text, we saw the reverse is true: for the full Hall crystals, even though the UV translations are magnetic and thus do not commute, the phonons appear as if there are commuting translations. The purpose of this appendix is to place a number of known physical results into the language and framework of this paper, which is a natural and transparent setting to understand several of them.

To derive the dynamical properties of skyrmions, we first compute
the Berry phase associated with a moving skyrmion. This section is a slight modification of the treatment in Ref.~\cite{stone1996magnus}, emphasizing the role played by the underlying lattice.

Consider a system of spin-$s$ particles in a square lattice on a torus with spacing $a$. For definiteness, take the particles at positions $\v{r}_{j} = a(n_j,m_j)$ for integers $n_j, m_j$, and let the periods of the rectangular torus be $L_x = N_x a$ and $L_y = N_y a$. Physically this is a model for an ionic crystal, a natural setting for a skyrmionic texture to develop.

We consider spin coherent states in this Hilbert space, which can be written as $\otimes_j \ket{\check{n}_j}$
where $\check{n}_j$ is a unit vector. The dynamics of these spin-coherent states can be mapped to the smooth continuum spin configuration $n(\v{r})$ by identifying
\begin{equation}
    n(\v{r}) \longleftrightarrow \otimes_j \ket{\check{n}_j = n(\v{r}_j)},
\end{equation}
where $\v{r}_j$ is the coordinate at site $j$. We can define a skyrmionic texture in our discretized Hilbert space by those corresponding to continuum $n(\v{r})$ with skyrmionic texture. As long as $n(\v{r})$ is slowly-varying, lattice configurations have a well-defined skyrmion winding number
\begin{equation}
    \mathrm{C}\Big[\otimes_j \ket{\check{n}_j = n(\v{r}_j)}\Big]
    = \frac{1}{4\pi} \int d^2\v{r} \; \check{n}(\v{r}) \cdot \partial_x \check{n}(\v{r}) \times \partial_y \check{n}(\v{r}) \in \mathbb{Z}.
\end{equation}

Similar to the main text, we now compute the Berry phase associated with a trajectory of a displaced skyrmion. We will define the trajectory of the quantum state to be generated by a time dependent texture $n(\v{r}, t)$ with period $T$: $n(\v{r}, T) = n(\v{r}, 0)$.
This corresponds to a parametrized set of states $\otimes_j \ket{n(\v{r}_j, t)}$. The Berry phase associated with this process is well-defined in the discretized Hilbert space, and is given by
\begin{equation}
    \varphi = \int_0^T dt \sum_j \bra{n(\v{r}_j, t)}i\partial_t\ket{n(\v{r}_j, t)} = s \sum_j \Omega_j,
\end{equation}
where $\Omega_j$ is the solid angle enclosed by the path $n(\v r_j, t)$ over one period.

Now consider, specifically, a trajectory of the form $n(\v r - \v u(t))$, where $n$ is a skyrmionic texture and $\v u(t)$ is the ``center-of-mass'' motion of the skyrmion. Then each spin undergoes a time evolution according to the local skyrmionic spin texture:
\begin{equation}
    \varphi_j = \int_0^T dt  
    \Braket{n(\v{r}_j - \v{u}(t))|i\partial_t|n(\v{r}_j -\v{u}(t))} = s A \lambda_j, \quad \lambda_j = \check{n}(\v{r}_j) \cdot \partial_x \check{n}(\v{r}_j) \times \partial_y \check{n}(\v{r}_j),
\end{equation}
where $A$ is the signed area associated with the path $\v{u}(t)$, and $\lambda_j$ is the skyrmion density at location $\v{r}_j$. The skyrmion density is quantized to be
\begin{equation}
     a^2 \sum_j \lambda_j = 4\pi C,
\end{equation}
where $C$ is the winding number of the skyrmion. Combining these results, we find the associated Berry phase to be
\begin{equation}
    s \sum_j \Omega_j = s A \sum_j \lambda_j =  \frac{4\pi sA}{a^2} C.
\end{equation}
We see that the Berry phase is proportional to the skyrmion number.

Let us now consider the effective action of the skyrmion motion. The only degree of freedom is given by the center of mass position $\v{u}$ of the skyrmion. The action to the quadratic order in time derivatives and $u$ is
\begin{equation}
    S = \frac{1}{2}\int dt \; (m \v{\dot{u}}^2 + \beta \v{u} \times \v{\dot{u}}).
    \label{eq:skyrmion_action}
\end{equation}
The Berry phase associated with a loop of area $A$ is then given by
\begin{equation}
    \varphi = \beta A.
\end{equation}
Comparing IR and UV computations, we find
\begin{equation}
    \beta = \frac{4\pi s}{a^2} C.
    \label{eq:skyrmion_berry_phase}
\end{equation}

Let us now consider this from the point of view of emanant symmetries. On the UV side, microscopic translations shift $\v{u}$ by $(a, 0)$ and $(0, a)$. Since these translations commute, the Berry phase from the loop around a unit cell (with area $a^2$) must be trivial:
\begin{equation}
    \beta a^2 = 2\pi m \implies \beta = \frac{2\pi}{a^2} m, 
\end{equation}
for some $m\in \Z$. Since $4\pi s \equiv 0 \pmod{2\pi}$, the Berry phase obtained in Eq.~\eqref{eq:skyrmion_berry_phase} is consistent with the emanant symmetry constraint.

We note that the action Eq.~\eqref{eq:skyrmion_action} is equivalent to that of a particle in a nonzero magnetic field. As a result, skyrmions feel Magnus force --- the counterpart of the Lorentz force on charged particles. This mixes longitudinal and transversal modes, resulting in anomalous dispersion for Skyrmion crystals~\cite{zang2011dynamics}.

We note that the original treatment in Ref.~\cite{stone1996magnus} was set in the lowest Landau level. In this case, instead of the ionic lattice, the ghost lattice discretizes the translation symmetry and fix the topological response.

If we quantize the skyrmion action Eq.~\eqref{eq:skyrmion_action}, we see that the momentum operators $\hat{P}_i$ no longer commute:
\begin{equation}
    [\hat{P}_x, \hat{P}_y] = i \frac{4\pi s}{a^2} C.
\end{equation}
The non-commutativity was discussed in detail from the field theory perspective in Ref.~\cite{watanabe2014noncommuting}.

In the analysis above, we focused on the simplified action of the form Eq.~\eqref{eq:skyrmion_action}, that only keeps the center of mass degree of freedom. Using coherent state path integrals, one can instead build a non-linear sigma model for the ferromagnet:
\begin{equation}
    \mathcal{L} = s \dot{\phi}(\cos\theta -1) - \frac{J}{2} \partial_i \check{n} \cdot \partial_i \check{n},
    \end{equation}
where the first term corresponds to the Berry phase term. Such a field theory does not admit a well-defined momentum operator, unless one fixes a topological sector; instead one must deal with discrete translation operators~\cite{Haldane1986}. These discrete translation operators exactly correspond to lattice translation operators in the UV, as was anticipated in the emanant symmetry discussion.
Various aspects of this phenomena have been studied in Refs.~\cite{Balakrishnan1985,Haldane1986,Volovik1987,Papanicolaou1991,Floratos1992,Banerjee1995,Nair2004,watanabe2014noncommuting,Tchernyshyov2015,Dasgupta2018,Di2021,Brauner2024, seiberg, seiberg2025anomalouscontinuoustranslations}. 

\section{Constraints on crystalline current from thermodynamics}
In this Appendix, we show that the Streda formula and adiabacity implies $\v{j} = e(\rho - C\phi) \v{v}$ \textit{up to a transversal contribution.}\footnote{We thank Bert Halperin for walking us through this argument.} Consider an adiabatic dynamical process corresponding to crystalline configuration $\v{u}(\v{r}, t)$. It has spatially dependent unit cell size $A$ given by
\begin{equation}
    \nabla \cdot \v{u}(x, t) = 1 - \frac{\delta A(x, t)}{{A_0}},
\end{equation}
where $A_0$ is the average size of the unit cell.
On the other hand, Streda formula and adiabaticity suggests the local charge density to be
\begin{equation}
    \rho(\v{r}, t) = \frac{s}{A} + C\phi \implies \delta \rho = - \frac{s}{{A_0}} \frac{\delta A (x, t)}{{A_0}}
\end{equation}
with \textit{spatially uniform} $s$. Combining these together, we get
\begin{equation}
    \nabla \cdot \v{u}(x, t) = 1 + \frac{{A_0}}{s}\delta\rho(x, t)
\end{equation}
From continuity equation,
\begin{equation}
\frac{\partial \delta \rho}{\partial t} - \nabla \cdot \v{j} = 0,
\end{equation}
we can solve for $\v{j}$ assuming it is linear in $\v{u}$:
\begin{equation}
    \v{j} = \frac{s}{{A_0}} \partial_t \v{u} + \alpha \nabla \times \partial_t \v{u} = (\rho - C\phi)\v{v} + \alpha \nabla \times \v{v}.
\end{equation}
Therefore, $\v{j}$ is determined \textit{up to} a transversal term. In the absence of spatial symmetries such as reflection, this transversal term cannot be ruled out. Our fully microscopic result, on the other hand, does not suffer from such a problem.

\end{document}